\documentclass[aps,prl,reprint,groupedaddress]{revtex4-2}
\usepackage{xparse}
\usepackage{physics}
\usepackage{bm}
\usepackage{hyperref}
\usepackage{siunitx}
\usepackage{graphicx}
\usepackage{titlesec}

\titleformat{\paragraph}[runin]
  {\normalfont\itshape}{\theparagraph.}{.5em}{}[.---]
\titlespacing{\paragraph}
  {\parindent}{1.5ex plus .1ex minus .2ex}{0pt}

\newcommand{\kB}{\ensuremath{k_\mathrm{B}}}
\newcommand{\kQTST}{\ensuremath{k_\mathrm{QTST}}}

\newcommand{\nbath}{\ensuremath{n_\mathrm{bath}}}

\newcommand{\Vext}{\ensuremath{V^\mathrm{ext}}}
\newcommand{\nmVsys}{\ensuremath{\nm{V}_{N}^{\mathrm{sys}}}}
\newcommand{\nmVnorm}{\ensuremath{\nm{V}_{N}^{\mathrm{ren}}}}
\DeclareSIUnit\hartree{\ensuremath{\mathit{E}}_h}
\DeclareSIUnit\angstrom{\ensuremath{\mathrm{\AA}}}
\DeclareSIUnit\wn{\ensuremath{\mathrm{cm}^{-1}}}
\NewDocumentCommand{\eu}{ !o }{ 
	\IfNoValueTF{#1}
		{\ensuremath{\mathrm{e}}}
		{\ensuremath{\mathrm{e}^{#1}}}
}
\NewDocumentCommand{\rpH}{ !o }{ 
	\IfNoValueTF{#1}
		{\ensuremath{H_N}}
		{\ensuremath{H_{#1}}}
}
\NewDocumentCommand{\nmH}{ !o }{ 
	\IfNoValueTF{#1}
		{\ensuremath{\widetilde{H}_N}}
		{\ensuremath{\widetilde{H}_{#1}}}
}
\NewDocumentCommand{\rpS}{ o }{ 
	\IfNoValueTF{#1}
		{\ensuremath{S_N}}
		{\ensuremath{S_{#1}}}
}
\NewDocumentCommand{\nmS}{ o }{ 
	\IfNoValueTF{#1}
		{\ensuremath{\widetilde{S}_N}}
		{\ensuremath{\widetilde{S}_{#1}}}
}
\NewDocumentCommand{\nm}{ m !o }{
	\IfNoValueTF{#2}
		{\ensuremath{\widetilde{#1}}}
		{\ensuremath{\widetilde{#1}^{(#2)}}}
}
\NewDocumentCommand{\xnm}{ !o }{
	\IfNoValueTF{#1}
		{\nm{q}}
		{\nm{q}[#1]}
}
\NewDocumentCommand{\wnm}{ !o }{
	\IfNoValueTF{#1}
		{\widetilde{\omega}}
		{\widetilde{\omega}_{#1}}
}
\NewDocumentCommand{\vbxnm}{ !o }{
	\IfNoValueTF{#1}
		{\nm{\vb{q}}}
		{\nm{\vb{q}}[#1]}
}
\NewDocumentCommand{\pnm}{ !o }{
	\IfNoValueTF{#1}
		{\nm{p}}
		{\nm{p}[#1]}
}
\NewDocumentCommand{\kernel}{ !o }{
	\IfNoValueTF{#1}
		{\ensuremath{\eta}}
		{\ensuremath{\eta^{(#1)}}}
}
\NewDocumentCommand{\resolvent}{ !o }{
	\IfNoValueTF{#1}
		{\ensuremath{K}}
		{\ensuremath{K^{(#1)}}}
}

\newcommand{\Eqn}[1]{Equation~\eqref{eq:#1}}
\newcommand{\eqn}[1]{Eq.~\eqref{eq:#1}}

\newcommand{\refx}[1]{Ref.~\citenum{#1}}
\newcommand{\fig}[1]{Fig.~\ref{fig:#1}}

\begin{document}

\title{Non-Markovian Effects in Quantum Rate Calculations of \\%
Hydrogen Diffusion with Electronic Friction}

\author{George Trenins}
\email{george.trenins@mpsd.mpg.de}
\author{Mariana Rossi}
\affiliation{MPI for the Structure and Dynamics of Matter, Luruper Chaussee 149, 22761 Hamburg, Germany}
\date{\today}

\begin{abstract}
	We address the challenge of incorporating non-Markovian electronic friction effects in quantum-mechanical approximations of dynamical observables. A generalized Langevin equation (GLE) is formulated for ring-polymer molecular dynamics (RPMD) rate calculations, which combines electronic friction with a description of nuclear quantum effects (NQEs) for adsorbates on metal surfaces. An efficient propagation algorithm is introduced that captures both the spatial dependence of friction strength and non-Markovian frictional memory. This framework is applied to a model of hydrogen diffusing on Cu(111) derived from \emph{ab~initio} density functional theory (DFT) calculations, revealing significant alterations in rate constants and tunnelling crossover temperatures due to non-Markovian effects. Our findings explain why previous classical molecular dynamics simulations with Markovian friction showed unexpectedly good agreement with experiment, highlighting the critical role of non-Markovian effects in first-principles atomistic simulations.
\end{abstract}

\maketitle

\paragraph*{Introduction}

Dissipative forces, arising from dynamical coarse-graining, play a crucial role in modelling multi-timescale systems~\cite{klippensteinIntroducingMemory2021}. A wide range of phenomena, including charge-transfer reactions, vibrational relaxation, and interfacial dynamics, can be effectively described as dissipative rate processes~\cite{Garg1985a,antoniouQuantumProton1999,tuckermanVibrationalRelaxation1993,polleyStatisticalMechanics2024a}. In particular, the dynamics of molecules on metal surfaces, influenced by electron--nuclear nonadiabatic coupling, are well-captured by electronic friction models~\cite{head-gordonMolecularDynamics1995,askerkaRoleTensorial2016,douPerspectiveHow2018}. While classical molecular dynamics with electronic friction (MDEF) has been successfully applied to problems such as vibrational relaxation~\cite{maurerInitioTensorial2016}, dissociative chemisorption~\cite{spieringOrbitalDependentElectronic2019}, and molecule-surface scattering~\cite{spieringTestingElectronic2018,gardnerAssessingMixed2023a}, it is limited by two significant shortcomings: the neglect of NQEs and the prevalent assumption of Markovian friction. For low-dimensional systems, these can be included exactly using the wavefunction~\cite{meyerMultidimensionalQuantum2009,bridgeQuantumRates2024} or reduced density-matrix~\cite{tanimuraNumericallyExact2020,Topaler1994,kunduPathSumFortran2023} formalism.

Imaginary-time path integral methods~\cite{feynman2010quantum} approximate quantum dynamical properties, but are readily applicable to fully atomistic simulations, otherwise intractable for exact approaches. They are effective at incorporating NQEs into rate calculations, with semiclassical instantons~\cite{Richardson2018} and ring-polymer molecular dynamics (RPMD)~\cite{Suleimanov2016} being standout approaches. Their strength lies in the ability to capture zero-point energy effects and incoherent quantum tunnelling in complex systems of many anharmonically interacting degrees of freedom, by harnessing the isomorphism between imaginary-time path integrals and the canonical partition function of an extended classical system~\cite{Chandler1981}. This enables the inclusion of NQEs at the computational cost of a geometry optimization or a classical MD simulation in an extended phase space.

A semiclassical instanton rate theory incorporating first-principles electronic friction was developed  by Litman and co-workers~\cite{litmanDissipativeTunneling2022,litmanDissipativeTunneling2022a}. The theory accurately describes the tunnelling-suppressing effects of friction, but is not suited, \emph{e.g.,} to surface dynamics in the low-friction regime, which is rate-limited by energy diffusion~\cite{Pollak1989,nitzanChemicalDynamics2006}. Instanton theory struggles to accurately capture this regime~\cite{pollakInstantonbasedKramers2024}, adding to its inherent challenges in describing low-barrier systems and reactions above the tunnelling crossover temperature~\cite{lawrenceSemiclassicalInstanton2024a}. 

Unlike instanton theory, RPMD does not suffer from these limitations and performs well across a wide range of friction strengths, as long as coherent nuclear tunnelling remains negligible~\cite{Richardson2009,fiechterHowQuantum2023,bridgeQuantumRates2024}.
For model systems, benchmarking against the multi-configuration time-dependent Hartree (MCTDH) method showed that RPMD can capture NQEs related to the spatial variation of the friction forces across a wide range of friction strengths~\cite{bridgeQuantumRates2024}. 
Recently, Bi and Dou~\cite{biElectronicFriction2024} performed RPMD simulations applying frictional forces to the ring-polymer centroids and assuming Markovian friction. Under these assumptions, the dissipative dynamics could be propagated efficiently using a modified version of the path-integral Langevin equation propagation algorithm~\cite{Ceriotti2010}. The restriction of frictional forces to the centroids, however, means that the suppression of the tunnelling probability is not captured~\cite{litmanDissipativeTunneling2022,litmanDissipativeTunneling2022a}. Furthermore, the description of non-Markovian effects is acknowledged as an outstanding challenge by the authors, highlighting the need for a more comprehensive solution.

Achieving a rigorous and, at the same time, efficient formulation of RPMD for dissipative systems thus remains an unsolved problem~\cite{bridgeQuantumRates2024,biElectronicFriction2024}.
Previously, Lawrence \emph{et al.}~\cite{Lawrence2019a} formulated a path-integral GLE for position-independent friction. Here, we generalize the derivation to include position-dependence of the friction and adopt the more efficient auxiliary variable propagation method~\cite{klippensteinIntroducingMemory2021}. This technique, explored by Ceriotti and co-workers for enhancing sampling in classical and path-integral MD~\cite{Ceriotti2010,ceriottiNuclearQuantum2009,ceriottiColoredNoiseThermostats2010,Ceriotti2012}, is repurposed to simulate the dissipative dynamics driven by an \emph{ab~initio} electronic memory--friction kernel. Using this implementation, we analyse a model of hydrogen diffusion on Cu(111), for which earlier classical MD simulations by Gu~\emph{et~al.}~\cite{guShortLongTime2022} showed close agreement with experimental measurements at \SI{200}{K}~\cite{townsendDiffusionLight2017}. This is unexpected given the significance of zero-point energy and shallow tunnelling in such systems around room temperature~\cite{caoDiffusionHydrogen1997,firminoDiffusionRates2014}. Our findings reveal that non-Markovian effects can explain this anomalous agreement.

\paragraph*{Theory} We begin by considering the classical Hamiltonian of an ($f+1$)-dimensional system,
\begin{equation}
	\label{eq:cl-ham}
    H(\vb{p}, \vb{q}) = \sum_{\nu=0}^{f}
        \frac{p_{\nu}^{2}}{2 m_{\nu}} + V(\vb{q}),
\end{equation}
where  $\vb{p} = \qty(p_0, \ldots,  p_f)$ and $\vb{q} = \qty(q_0,  \ldots,  q_f)$ denote the Cartesian momentum and position coordinates, respectively, and $V$ is the potential energy. To construct the RPMD Hamiltonian~\cite{Craig2004}, we introduce $N$ replicas of the original system, also referred to as beads, which are connected into a ring polymer by harmonic springs. The RPMD Hamiltonian is given by
\begin{equation}
	\label{eq:rp-ham}
    \rpH(\bm{p},\bm{q}) = \sum_{l=0}^{N-1} H\qty\big(\vb{p}^{(l)}\!, \vb{q}^{(l)}) + 
    \sum_{\nu=0}^{f} \rpS\qty\big(\vb{q}_{\nu}),
\end{equation}
where  $\vb{q}^{(l)} = \qty(\smash{q_0^{(l)},  \ldots,  q_f^{(l)}})$ represents the position coordinates of the \mbox{$l$-th} replica and
$\vb{q}_{\nu} = \qty(\smash{q_{\nu}^{(0)},  \ldots,  q^{(l-1)}_{\nu}})$ denotes the \mbox{$\nu$-th} components of the position coordinates of all replicas. The harmonic spring term,
\begin{equation}
    \label{eq:spring}
    \rpS\qty\big(\vb{q}_{\nu}) = \sum_{l=0}^{N-1} \frac{m_{\nu} \omega_N^2}{2} 
    \qty(q_{\nu}^{(l+1)} \! - q_{\nu}^{(l)})^2
\end{equation}
depends on temperature $T$ through the harmonic frequency $\omega_N = N/\beta \hbar$, where $\beta = 1/\kB T$. 

In the limit of large $N$, typically around $\order{10}$--$\order{100}$,
the classical canonical distribution of RPMD position coordinates converges to the exact quantum Boltzmann distribution. 
This property, along with others discussed in Refs.~\citenum{Richardson2009},
\citenum{Hele2013}, and \citenum{Hele2015a}, ensures that RPMD reaction rates provide a good approximation to quantum thermal rates in the absence of substantial quantum coherence~\cite{Craig2005a,Craig2005}. RPMD rates can be computed using the Bennett--Chandler approach~\cite{Collepardo-Guevara2008}, which expresses the thermal rate as the product 
\begin{equation}
	\label{eq:rpmd-rate}
	k(T) = \kappa(t_p) \kQTST(T),
\end{equation}
where $\kQTST(T)$ is the quantum transition-state theory rate~\cite{Hele2013,Althorpe2013} and 
$\kappa(t_p)$ is the dynamical transmission coefficient. The latter is calculated as the value of the flux-side correlation function at a plateau time $t_p$~\cite{Craig2005a,Craig2005}. \Eqn{rpmd-rate} relies on a separation of timescales between the recrossing dynamics and the reaction~\cite{Craig2007}. For low-barrier systems, where this assumption may not hold, a modified expression applies, given in the Supplemental Material (SM)~\footnotemark[1].

\paragraph*{Friction}
To incorporate electronic friction into RPMD, we consider the potential energy function
\begin{equation}
	\label{eq:V-sysbath}
    V(\vb{q}) = \Vext(Q) +  \sum_{\nu=1}^{\nbath} \frac{m \omega_{\nu}^{2}}{2}\qty[
        q_{\nu} - \frac{c_{\nu} F(Q)}{m \omega_{\nu}^{2}}
    ]^{2},
\end{equation}
where $(P,Q) \equiv (p_0,q_0)$ denotes the nuclear coordinates---here taken to be one-dimensional---of a species with mass~$m$. $\Vext(Q)$ denotes the Born--Oppenheimer potential energy surface, and the sum over $\nu$ represents the coupling to a bath of harmonic oscillators. 
This is a general mapping for dissipative systems whose dynamics in the classical limit are described by the GLE in Eq.~(S5)~\cite{Note1}. It does not assume that the fine-grained dynamics of the dissipative environment is literally harmonic (see Appendix~C of \refx{Caldeira1983}).
The coupling coefficients $c_{\nu}$ and the bath frequencies $\omega_{\nu}$ are encoded in the spectral density
\begin{equation}
	J(\omega) = \frac{\pi}{2} \sum_{\nu=1}^{\nbath}
	\frac{c_{\nu}^{2}}{m \omega_{\nu}} \delta(\omega - \omega_{\nu})
\end{equation}
or, equivalently, the spectrum
\begin{equation}
	\Lambda(\omega) = J(\omega) / \omega.
\end{equation}
The friction kernel $\eta(Q, t, t')$ is proportional to the cosine transform of $\Lambda(\omega)$, as defined in the SM~\cite{Note1}.
Any physical friction can be accurately represented using a sufficiently large $n_{\text{bath}}$. In this study, we obtain the spectrum $\Lambda(\omega)$ and the interaction potential $F(Q)$ from \emph{ab~initio} DFT calculations, identifying $F(Q)^2 \Lambda(\omega)$ with Eq.~(14) of 
Ref.~\cite{boxInitioCalculation2023}.

The system-bath potential in \eqn{V-sysbath} can be constructed by a harmonic discretization of the spectral density~\cite{waltersDirectDetermination2017} and ring-polymerized as in \eqn{rp-ham}. This provides an ``explicit'' representation of the dissipative environment in RPMD, employed by Bridge \textit{et~al.} in a previous study comparing dissipative RPMD rates to exact quantum benchmarks~\cite{bridgeQuantumRates2024}. The explicit representation requires many bath modes to converge the dissipative dynamics, introducing a prohibitive computational overhead.

\paragraph*{Path-Integral GLE}
To reduce the overhead, we solve for the motion of the harmonic bath modes analytically, as done in Ref.~\citenum{Lawrence2019a} for position-independent friction, \emph{i.e.,} $F(Q) = Q$. In the SM~\cite{Note1}, we present all equations of motion for a general $F(Q)$, which encompasses position-dependent friction.
We formulate the path-integral GLE in ring-polymer ``normal-mode'' coordinates, $\nm{\vb{Q}} = (\nm{Q}^{(0)}\! , \nm{Q}^{(\pm 1)}\! , \nm{Q}^{(\pm 2)}\! , \ldots )$. These are the normal modes of a free ring-polymer, which are known analytically. Their definition is given in the SM~\footnotemark[1] along with the derivation of the ring-polymer GLE.

The dissipative bath renormalizes the ring-polymer potential, leading to
\begin{subequations}
	\label{eq:bob}
	\begin{align}
		\label{eq:rV-renorm}
		\nmVnorm(\nm{\vb{Q}}) & = \nmVsys(\nm{\vb{Q}}) + \frac{1}{2} \sum_{\smash[b]{n}} \alpha^{(n)} \qty\big[\nm{F}^{(n)}]^2, \\
		\label{eq:alpha-renorm}
		\alpha^{(n)} & =  \frac{2}{\pi} \int_{0}^{\infty} 
		\frac{\Lambda(\omega)  \wnm[n]^2}{\omega^2 + \wnm[n]^2} \dd{\omega},
	\end{align}
\end{subequations}
where $\nmVsys$ is the potential of an $N$-bead ring polymer in the absence of friction, $\nm{F}^{(n)}$ are normal-mode interaction potentials, related to $\qty{F(Q^{(0)}),\,F(Q^{(1)}),\ldots,F(Q^{(N-1)})}$ by a linear transformation, and
\begin{equation}
	\wnm[n] = \frac{2N}{\beta\hbar} \sin(\frac{\pi\abs{n}}{N})
\end{equation}
are the harmonic frequencies associated with normal modes $\nm{Q}[\pm n]$.
\Eqn{bob} is a reformulation of Eq.~(34) in Ref.~\cite{litmanDissipativeTunneling2022}.
It accounts for the suppression of tunnelling probability and is the only modification needed to obtain the QTST rate in the presence of friction.

The dynamical transmission coefficient is further influenced by the dissipative and stochastic forces acting on the normal-mode momenta $\nm{P}[n]$. These forces are described by the GLE
\begin{align}
	\label{eq:rpmd-gle}
	\dv{\nm{P}[n]}{t} & = - \pdv{\nmVnorm}{\nm{Q}[n]} - \! \int^{t} \!\! \sum_{n'} \eta^{(n,n')}(t,t'; \nm{\vb{Q}}) \, \frac{\nm{P}[n'] (t')}{m} \dd{t'} \nonumber \\ 
	& + \sum_{n'} \pdv{\nm{F}[n'](t)}{\nm{Q}[n]} \zeta^{(n')}(t),
\end{align}
where the ring-polymer friction tensor is given by
\begin{equation}
	\eta^{(n,n')}(t,t'; \nm{\vb{Q}}) = 
	\sum_{n''}\pdv{\nm{F}[n''](t)}{\nm{Q}[n]} K^{(n'')}(t-t') \pdv{\nm{F}[n''](t')}{\nm{Q}[n']},
\end{equation}
and the ``resolvent'' associated with $\nm{P}[n]$ is
\begin{equation}
	\label{eq:rpmd-resolvent}
	\resolvent[n](t) = \frac{2}{\pi} \int_{0}^{\infty} \!
	\frac{\Lambda(\omega)  \omega^2}{\omega^2 + \wnm[n]^2} \cos(
		t \sqrt{\omega^2 + \wnm[n]^2}
	) \dd{\omega}.
\end{equation}
The stochastic forces $\zeta^{(n)}$ satisfy the second fluctuation-dissipation theorem~\cite{kuboFluctuationdissipationTheorem1966}
\begin{equation}
	\expval*{\zeta^{(n)}(t) \mkern1mu \zeta^{(n)}(t')} = \kB T K^{(n)}(t-t').
\end{equation}
For simplicity, all expressions are given for a one-dimensional system but can be straightforwardly generalized to the multidimensional case.

For the centroid coordinate $\nm{P}[0]$, the normal-mode frequency $\wnm[0] = 0$, so that the centroid resolvent $K^{(0)}(t)$ is identical to the resolvent of the classical GLE. For a constant friction spectrum [$\Lambda(\omega) = \eta_0$] the centroid resolvent produces a Markovian friction, $K^{(0)}(t) \propto \delta(t)$. The other resolvents have a more complicated time dependence, even for constant $\Lambda(\omega)$~\cite{Lawrence2019a}, and will always generate non-Markovian friction. To propagate the dissipative dynamics, we map the non-Markovian GLE onto a system of 
multivariate Ornstein--Uhlenbeck equations~\cite{gardinerHandbookStochastic1985}. The mapping consists in coupling $n_{\text{aux}}$ pairs of auxiliary dynamical variables to every momentum coordinate. The coupling parameters are fitted to reproduce the dissipative and random forces in \eqn{rpmd-gle} (see SM~\footnotemark[1] for details of the parametrization and propagation). This approach generalizes the path-integral stochastic thermostatting algorithm~\cite{ceriottiNuclearQuantum2009,ceriottiColoredNoiseThermostats2010,Ceriotti2012}, and requires only a few auxiliary variables 
($n_{\text{aux}} \ll n_{\text{bath}}$) to converge the dissipative dynamics, significantly reducing the computational overhead.

\begin{figure}[t]
	\includegraphics{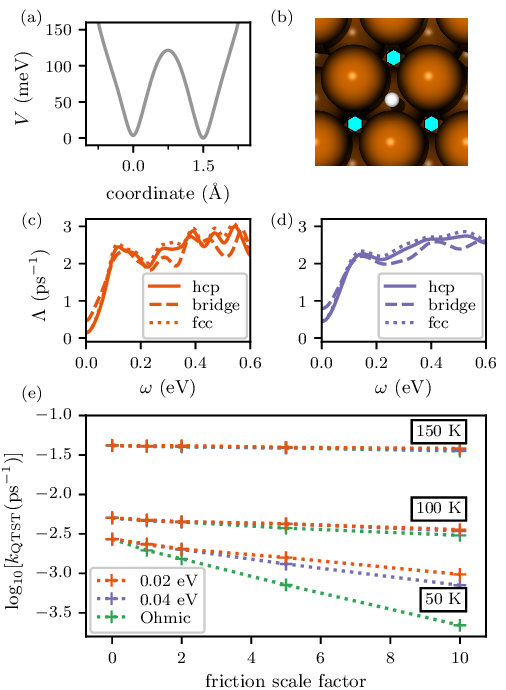}%
	\caption{%
	(a)~Potential energy along the reaction coordinate connecting the \emph{hcp} site of Cu(111) at \SI{0}{\AA} and the \emph{fcc} site at \SI{1.5}{\AA}, computed using DFT with the PBE exchange--correlation functional and Tkatchenko--Scheffler
	screened van der Waals correction (see SM~\cite{Note1}).
	(b)~Hydrogen atom at the \emph{hcp} site of Cu(111), with the three neighbouring \emph{fcc} sites marked by blue hexagons. 
	(c)~Spectra of the electronic-friction tensor projected onto the reaction coordinate and convolved with a $\sigma = \SI{0.02}{eV}$ Gaussian window. The solid, dashed, and dotted lines show spectra at the reactant minimum (\emph{hcp}), transition state (bridge) and product minimum (\emph{fcc}), respectively.
	(d)~Same for $\sigma = \SI{0.04}{eV}$. 
	(e)~QTST rates for the potential in panel~(a), coupled to a bath with either the \emph{fcc} spectrum from panel~(c), panel~(d), or an Ohmic bath with $\eta_0 = \SI{2}{ps^{-1}}$. Rates are also shown for systems with rescaled friction, $\eta_{\text{scaled}}(Q, t, t') = \varsigma \eta(Q, t, t')$, where 
    $\varsigma$ is the scale factor.
	\label{fig:tst}}
\end{figure}

\paragraph*{Reaction Model}

To study the diffusion of atomic hydrogen on Cu(111), we mapped the potential connecting the \emph{hcp} and \emph{fcc} adsorption sites [\fig{tst}(a,b)]. We considered the effects of surface relaxation, basis set, and choice of functional, as reported in the SM~\cite{Note1}. Below, we show results for a substrate lattice that was fixed at the bare-slab equilibrium geometry. For the mapping, the hydrogen atom was constrained at equidistant points along the line connecting the two adsorption sites, optimizing the displacement perpendicular to the slab surface.  We calculated the potential energy using  DFT with the PBE exchange-correlation functional~\cite{perdewGeneralizedGradient1996} and the Tkatchenko--Scheffler screened \mbox{van der Waals} dispersion corrections~\cite{ruizDensityFunctionalTheory2012}, as implemented in FHI-aims~\cite{blumInitioMolecular2009} (see the SM~\footnotemark[1]). The electronic friction was computed from the electron-phonon coupling matrix elements, also obtained using FHI-aims~\cite{boxInitioCalculation2023}. Deviations of the interaction potential $F(Q)$ from linearity were negligible for this system, and all results presented below were derived from simulations using the linear approximation, $F(Q) = Q$. Dynamical simulations were performed with an in-house code~\cite{rpmd-gle-inhouse}, where the algorithms developed in this letter were implemented for one-dimensional potential energy surfaces.

\begin{figure}[b]
	\includegraphics{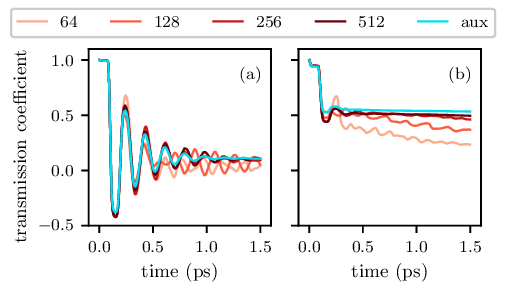}%
	\caption{%
	(a)~Classical ($N=1$) transmission coefficient [$\kappa(t)$ from \eqn{rpmd-rate}] at \SI{200}{K} for the \emph{fcc} friction spectrum in \fig{tst}(c) damped by  $w(\omega) = \eu[-\omega/\omega_c]$, $\omega_c = \SI{4000}{\wn}$. Calculations using harmonic bath discretization \cite{waltersDirectDetermination2017} are shown in shades of red (see legend for $n_{\text{bath}}$). Simulation results using the auxiliary variable propagator with $n_{\text{aux}} =5 $ are shown in cyan.
	(b)~Same for friction scaled by a factor of~10.
	\label{fig:kappa}}
\end{figure}

\paragraph*{QTST Rates}

In a typical calculation, the ``raw'' \emph{ab~initio} electronic friction spectrum is convolved 
with a Gaussian window of width $\sigma$ to extrapolate to the infinitely large $k$-grid limit. In order to approximate the friction as Markovian, $\sigma$ is set to be large, typically between $\SI{0.3}{eV}$ and $\SI{0.6}{eV}$. However, this washes out the spectral detail at the low energy scales relevant to surface diffusion. A narrower window [\fig{tst}(c) and~(d)] reveals that for hydrogen on Cu(111) electronic friction is super-Ohmic~\cite{weissQuantumDissipative2012} at low energies.
This feature has a significant impact on calculated reaction rates. The choice of $\sigma$ 
directly impacts the renormalized ring-polymer potential in \eqn{bob}, which modulates tunnelling probabilities and the computed QTST rates. 

Using the renormalized ring-polymer potential instead of the system-bath representation in \eqn{V-sysbath} greatly accelerated the convergence of QTST rate calculations. The acceleration is twofold: firstly, there are fewer degrees of freedom to propagate, so the computational cost of a single propagation step is reduced. Secondly, 
we are not explicitly sampling the thermal distribution of the harmonic bath modes, so that a given sampling accuracy is reached for a smaller total number of steps (see SM~\cite{Note1}). Simulation times in our study were reduced by a factor of 100 for some friction and temperature regimes.

To facilitate comparison with experimental results, all calculated QTST rates are multiplied by a factor of 3, in order to account for hops to all three equivalent neighbouring \emph{fcc} sites [\fig{tst}(b)].
For the \emph{ab~initio} friction, the relative differences between rates calculated with Ohmic and super-Ohmic profiles significantly increase when the  
friction strength is varied within the range spanned by typical metal substrates, as shown in \fig{tst}(e). Thus, we predict that non-Markovian effects will have a stronger impact on low-temperature QTST rates for certain surfaces, e.g., Ru(0001)~\cite{townsendDiffusionLight2017,gerritsElectronicFriction2020},

\paragraph*{Overall rates}
 
Above \SI{150}{K}, the impact of friction on the escape rates is largely determined by the transmission coefficient $\kappa(t_p)$. Calculating $\kappa$ for our system is especially challenging due to the weak damping exerted by the electronic friction. 
We still expect
RPMD to provide a good description of the quantum reaction rates, 
since quantum coherence should be negligible for low-frequency, low-barrier processes~\cite{fiechterHowQuantum2023a,bridgeQuantumRates2024}.
However, weak friction results in a flux-side correlation function that is slow to reach a plateau. If one were to simulate the dissipative dynamics using a harmonic system-bath mapping in \eqn{V-sysbath} and the discretization procedure in Ref.~\citenum{waltersDirectDetermination2017}, over 500 modes would be needed to converge the calculation [\fig{kappa}(a) and (b)].
Simulations using the method of auxiliary variables to propagate the GLE in \eqn{rpmd-gle} reach convergence with only $n_{\text{aux}} = 5$. 

\begin{figure}[t]
	\includegraphics{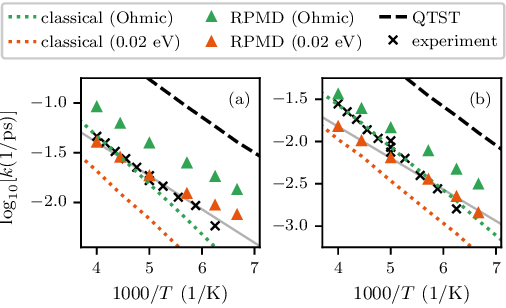}%
	\caption{%
	(a)~Hydrogen escape rates from the \emph{hcp} site of Cu(111) for the potential in \fig{tst}(a) compared against HeSE measurements~\cite{townsendDiffusionLight2017} (black crosses). The black dashed line is the QTST rate, which is essentially identical with the semiclassical Wolynes rates~\cite{wolynesQuantumTheory1981} in Ref.~\citenum{townsendDiffusionLight2017}.
	The grey line shows the linear Arrhenius fit to RPMD rates between 200 and \SI{250}{K}.
	(b)~Same for deuterium. 
 	\label{fig:rates}}
\end{figure}

As seen from \fig{rates}, the Markovian approximation leads to a substantial overestimation of the rate, 
which is limited by the efficiency of energy dissipation at this friction strength~\cite{Pollak1989,nitzanChemicalDynamics2006}.  The characteristic frequencies associated with the reaction coordinate are of $\order{\SI{50}{meV}}$, a range where Ohmic friction exhibits enhanced spectral density compared to the actual spectra in \fig{tst}(c).
Hence, Ohmic friction artificially enhances energy dissipation, leading to overestimated rates.

A comparison of the low-friction calculation results with the \emph{hcp}-site escape rates extracted from \mbox{helium-3} surface spin-echo (HeSE) measurements by Townsend and co-workers~\cite{townsendDiffusionLight2017} confirms that classical dynamics with Markovian friction provides a surprisingly good fit to the data, as noted previously~\cite{guShortLongTime2022}. However, a closer examination of the temperature dependence reveals the limitations of the classical model. Specifically, for hydrogen [\fig{rates}(a)], the experimental rates at low temperatures ($< \SI{200}{K}$) exceed the predictions of a classical Arrhenius curve (see also Fig.~S5 in the SM~\footnotemark[1]). This discrepancy arises from the contribution of activationless deep tunnelling, which becomes the dominant reaction mechanism below the crossover temperature (${\approx\SI{100}{K}}$). The onset of this effect is already apparent at the experimentally probed temperatures and cannot be captured by the classical model alone.

In contrast to hydrogen, the experimental rates for deuterium [\fig{rates}(b)] deviate from the linear Arrhenius trend in the opposite sense, falling slightly below the expected values at low temperatures. For this reaction, the crossover to deep tunnelling occurs around \SI{70}{K} and does not significantly impact the temperature range in \fig{rates}(b). However, the heavier mass of deuterium leads to RPMD rates converging on the classical predictions already at around \SI{250}{K}, due to the diminishing importance of zero-point energy and shallow tunnelling effects~\cite{bell2013tunnel,Richardson2016b}. Consequently, the RPMD data for $T > \SI{200}{K}$, when fitted to a straight line, create the appearance of a low-temperature rate ``suppression'', mirroring the experimental observation.

\paragraph*{Outlook}
In this letter, we described how non-Markovian friction effects can be incorporated into RPMD simulations using a GLE framework and an auxiliary-variable propagator algorithm. This method is readily applicable to multidimensional systems and it is generalizable to different types of frictional forces. It offers a practical tool for atomistic modelling, requiring only minor modifications to existing path-integral GLE infrastructure, such as that implemented in \mbox{i-PI}~\cite{litmanIPI302024}. 

By applying this approach to the problem of hydrogen diffusion on Cu(111), we could study the interplay between friction memory and NQEs in reaction rates. Our results revealed that friction memory can significantly influence reaction rates above the tunnelling crossover temperature, effectively masking NQEs. This provides a nuanced understanding of the previously reported agreement between classical molecular dynamics and experiment~\cite{guShortLongTime2022}, highlighting the importance of accurate electronic structure, NQEs, and dissipative dynamics for quantitative modelling. Furthermore, the auxiliary-variable GLE formulation is applicable to systems with strongly position-dependent friction, where we expect it to yield new physical insights~\cite{litmanDissipativeTunneling2022,litmanDissipativeTunneling2022a}. 
Combined with advances in embedding techniques~\cite{zhaoRevisitingUnderstanding2021,wenStrategiesObtain2024}, machine learning potentials~\cite{kulikRoadmapMachine2022} and electronic friction tensors~\cite{sachsMachineLearning2025}, this method will be a powerful approach for studying reactions in metallic environments.

\begin{acknowledgments}
\paragraph*{Acknowledgments}
We thank Yair Litman and Paolo Lazzaroni for our discussions of GLEs with position-dependent friction. We also thank Connor Box and Reinhard Maurer for their insightful advice on computing electronic friction. We are grateful to Paolo Lazzaroni for providing explicit system-bath benchmark RPMD rates. G.T. gratefully acknowledges a Research Fellowship from the Alexander von Humboldt Foundation.
\end{acknowledgments}

\footnotetext[1]{See Supplemental Material at [URL will be inserted by publisher] 
for details of the DFT calculations,
potential energy surface, ring-polymer GLE, auxiliary variable propagation algorithm, and details of 
the RPMD rate calculations. The Supplemental Material also includes Refs.~\cite{
stellaGeneralizedLangevin2014,
liuStructureEnergetics2013,
monkhorstSpecialPoints1976,
neugebauerAdsorbatesubstrateAdsorbateadsorbate1992,
zhaoNewLocal2006,
oudotReactionBarriers2024,
tuckermanStatisticalMechanics2010,
ceriottiPhD2010,
NumRep}.}

\end{document}


\title{Supplemental Material for: Non-Markovian Effects\\%
in Quantum Rate Calculations of
Hydrogen Diffusion with Electronic Friction}

\author{George Trenins}
\email{george.trenins@mpsd.mpg.de}
\author{Mariana Rossi}
\affiliation{MPI for the Structure and Dynamics of Matter, Luruper Chaussee 149, 22761 Hamburg, Germany}

\date{\today}

\maketitle
\onecolumngrid

\vspace*{-3em}

\section{DFT calculations}

Unless noted otherwise, all calculations were done in FHI-aims~\cite{blumInitioMolecular2009} using the PBE functional~\cite{perdewGeneralizedGradient1996}. The default \emph{light}
settings were used for the basis sets, and modified \emph{light} settings with double radial density were used for real-space integration grids. Optimized geometries described below were essentially identical when using \emph{tight} defaults.

The bulk structure of Cu was optimized using a primitive cell with a 16$\times$16$\times$16 $k$-grid (Monkhorst-Pack  grids are used throughout~\cite{monkhorstSpecialPoints1976}), constrained to the face-centred cubic (FCC) symmetry, converging to a conventional lattice parameter $a = \SI{3.631}{\AA}$. A~bare 1$\times$1$\times$5 Cu(111) slab with the bottom two layers fixed at the bulk geometry was placed in the middle of \SI{100}{\AA} cell and allowed to relax, now using a 16$\times$16$\times$1 $k$-grid. In this and all subsequent slab calculations we applied a dipole correction~\cite{neugebauerAdsorbatesubstrateAdsorbateadsorbate1992}. The interlayer spacings, going from the bulk to the surface, was calculated to be \SI{2.096}{\AA}, \SI{2.090}{\AA}, \SI{2.091}{\AA}, and \SI{2.078}{\AA} at equilibrium. 

The relaxed slab was tiled laterally, to form a 4$\times$4$\times$5 supercell, and a hydrogen atom was placed at $K=11$ equally spaced points along a straight line connecting a pair of neighbouring \emph{hcp} and \emph{fcc} adsorption sites. From here onwards, the PBE functional was augmented by a dispersion interaction correction, described by the screened vdW\textsuperscript{surf} model~\cite{ruizDensityFunctionalTheory2012} with coefficients for Cu 
taken from \refx{liuStructureEnergetics2013} and with Cu--Cu interactions excluded. Using a 4$\times$4$\times$1 $k$-grid, the $z$-coordinate of the hydrogen atom was allowed to relax, while holding all other coordinates fixed. The resulting geometries were used 
to run the convergence tests in \sec{sec:dft-convergence}, removing or adding layers of Cu atoms at the bulk geometry to the bottom of the slab as needed.

\subsection{Convergence tests%
\label{sec:dft-convergence}}

\begin{figure}[t]
	\includegraphics{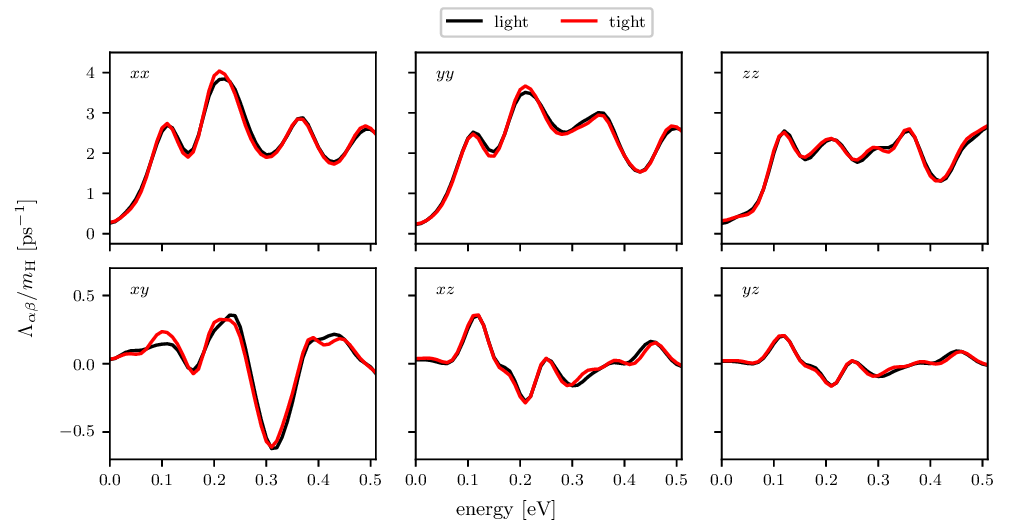}%
	\caption{%
	Spectra of the EFT elements calculated using a 10$\times$10$\times$1 $k$-grid for a 5-layer Cu(111) slab at the \emph{fcc} site using modified \emph{light} and \emph{tight} settings. The Gaussian broadening of the spectrum was set to $\sigma = \SI{0.02}{eV}$. Here, $m_{\text{H}} = \SI{1.007825}{\dalton}$ is the mass of a hydrogen-1 atom. 
	\label{fig:basis-conv}}
\end{figure}

\begin{figure}[b!]
	\includegraphics{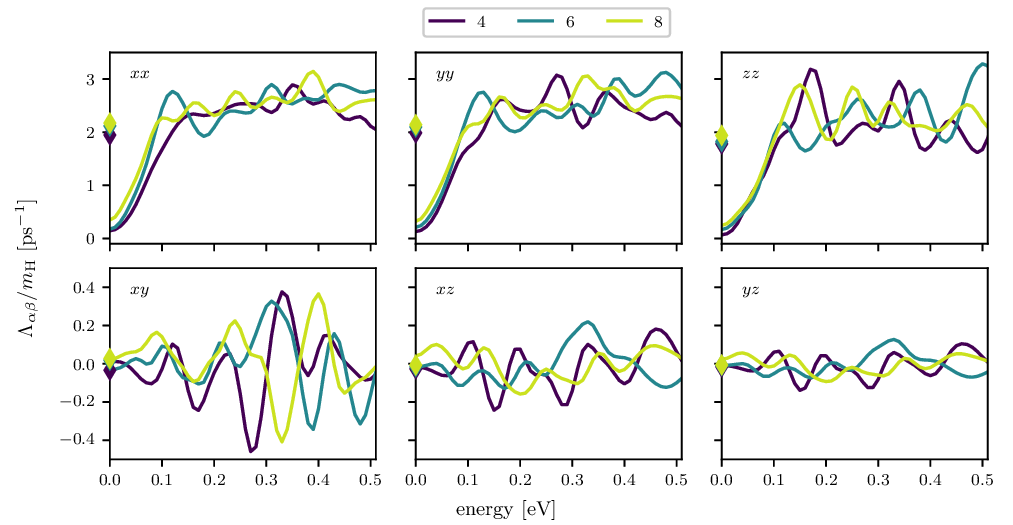}%
	\caption{Same as \fig{basis-conv}, except the modified \emph{light} settings are used throughout and the thickness of the metal slab is varied (number of metal atom layers given in the legend). The diamond markers indicate the components of the static friction tensor, computed as $\Lambda(0)_{\alpha\beta} / m_{\text{H}}$ for a heavily broadened spectrum ($\sigma = \SI{0.3}{eV}$).
	\label{fig:layer-conv}}
\end{figure}

We first tested the convergence of the hydrogen-atom electronic friction tensor (EFT) at the \emph{fcc} site 
with respect to the basis set and the number of metal layers. The EFT was computed from the raw electron--phonon coupling (EPC) matrix elements according to Eq.~(14) of \refx{boxInitioCalculation2023}, using a Gaussian kernel with a standard deviation $\sigma=\SI{0.02}{eV}$ to represent the delta functions.
The EPC matrix elements were computed using the expression proposed by Head-Gordon and Tully~\cite{head-gordonMolecularDynamics1995}, given in Eq.~(28) of \refx{boxInitioCalculation2023}.
Like the equilibrium geometries, the EFT spectra are converged with respect to the basis set and numerical quadrature grids (\fig{basis-conv}). Convergence with metal slab thickness is sufficient at 6~atom layers  (\fig{layer-conv}). 

To assess convergence with the size of the $k$-grid, we looked at projections of the EFT onto the reaction coordinate. Denoting with $\vb{q}_i$ the position of the hydrogen atom at the $i$-th grid point between the \emph{hcp} and \emph{fcc} sites, we defined the reaction coordinate as
\begin{equation}
    s_n = \sum_{i=1}^{n} \delta s_i \qqtext{where} \delta s_i = \norm{\vb{q}_i - \vb{q}_{i-1}} \qqtext{and} s_0 = 0.
\end{equation}
The tangent to the reaction path is computed using a generalized central finite difference that has an error of $\order*{\delta s^3}$,
\begin{equation}
\vb{t}_i = \qty[\delta s_{i+1} + \delta s_{i}]^{-1} \qty[
    \qty(\vb{q}_{i+1} - \vb{q}_{i}) \frac{\delta s_{i}}{\delta s_{i+1}} +
    \qty(\vb{q}_{i} - \vb{q}_{i-1}) \frac{\delta s_{i+1}}{\delta s_{i}}
].
\end{equation}
The values at the end-points (the \emph{hcp} and \emph{fcc} sites) are obtained by imposing mirror boundary conditions, amounting to $t_{z,0} = t_{z,K} = 0$, and
\begin{align}
	t_{\alpha,0} & = (\alpha_1 - \alpha_0) / \delta s_1, & 
	t_{\alpha,K} & = (\alpha_K - \alpha_{K-1}) / \delta s_K
\end{align}
for $\alpha \equiv x$ or $y$. 
The projection of the EFT onto the reaction coordinate at the point $s_n$ is then
\begin{equation}
    \Lambda(\omega; s_n) = \vb{t}_n \vdot \bm{\Lambda}(\omega; \vb{q}_n) \vdot \vb{t}_n / \norm{t_n}^2.
\end{equation}

\begin{figure}[t]
	\includegraphics{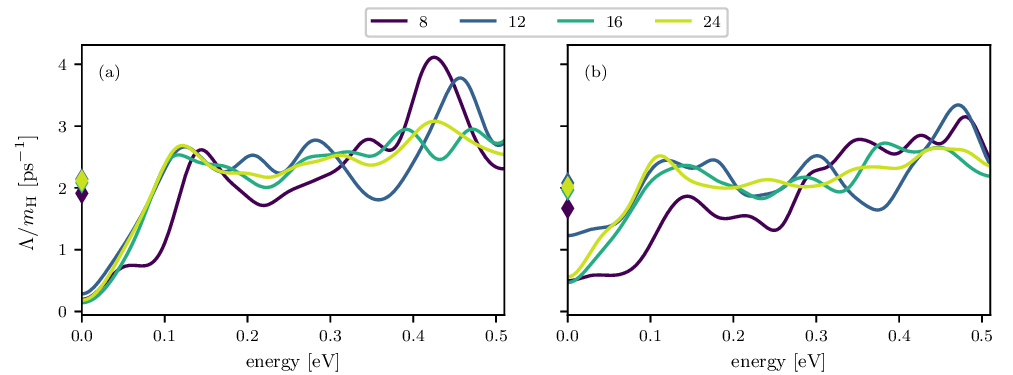}%
	\caption{Spectra of the EFT for a 6-layer Cu(111) slab, projected onto the reaction coordinate at (a) the \emph{fcc} site and (b) the bridge site. The calculation was performed using modified \emph{light} settings, a Gaussian spectral broadening of $\sigma = \SI{0.02}{eV}$, and $k$-grids of size 
	$n_k\mathord{\times}n_k\mathord{\times}1$, where $n_k$ is specified in the legend.
	The diamond markers indicate the values of static friction, computed as $\Lambda(0) / m_{\text{H}}$ for a heavily broadened spectrum ($\sigma = \SI{0.3}{eV}$).
	\label{fig:kpts-conv}}
\end{figure}

In \fig{kpts-conv} we show how the projected spectrum at the \emph{fcc} site and the bridge site (our transition state) changes upon increasing the size of the $k$-grid. We judge the spectra to be sufficiently well converged for a 16$\times$16$\times$1 grid. The potential energies along the reaction coordinate were converged already for a 12$\times$12$\times$1 grid.

\subsection{Effect of surface relaxation and choice of functional}

All simulation results in the main article are reported for potential energies computed for the frozen slab geometry using the PBE functional with modified light settings defined previously. Relaxing the surface geometry, using tighter settings or changing the exchange--correlation functional all produce qualitatively similar models, which does not affect our conclusions regarding the magnitude and importance of nuclear quantum effects and non-Markovian friction. However, the energy changes we observe have an appreciable effect on the rates, and careful benchmarking of the electronic structure will be necessary to reach quantitative agreement with experiment in full-dimensional calculations.

\begin{figure}[h!]
	\includegraphics{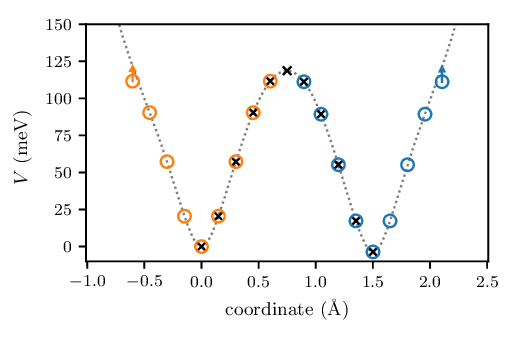}%
	\caption{The black crosses show the \emph{ab~initio} energies computed along the reaction coordinate between the \emph{hcp} site at \SI{0.0}{\AA} and the \emph{fcc} site at \SI{1.5}{\AA}. The energies marked with orange circles are obtained by mirroring the data about the bottom of the reactant (\emph{hcp}) well. The blue circles denote the same for the product (\emph{fcc}) well. A positive bias of \SI{10}{meV} is applied to the outer points (vertical arrows). A cubic spline is then fitted to obtain the model PES, shown with a dotted grey line.
	\label{fig:pes-build}}
\end{figure}

\begin{figure}[t]
	\includegraphics{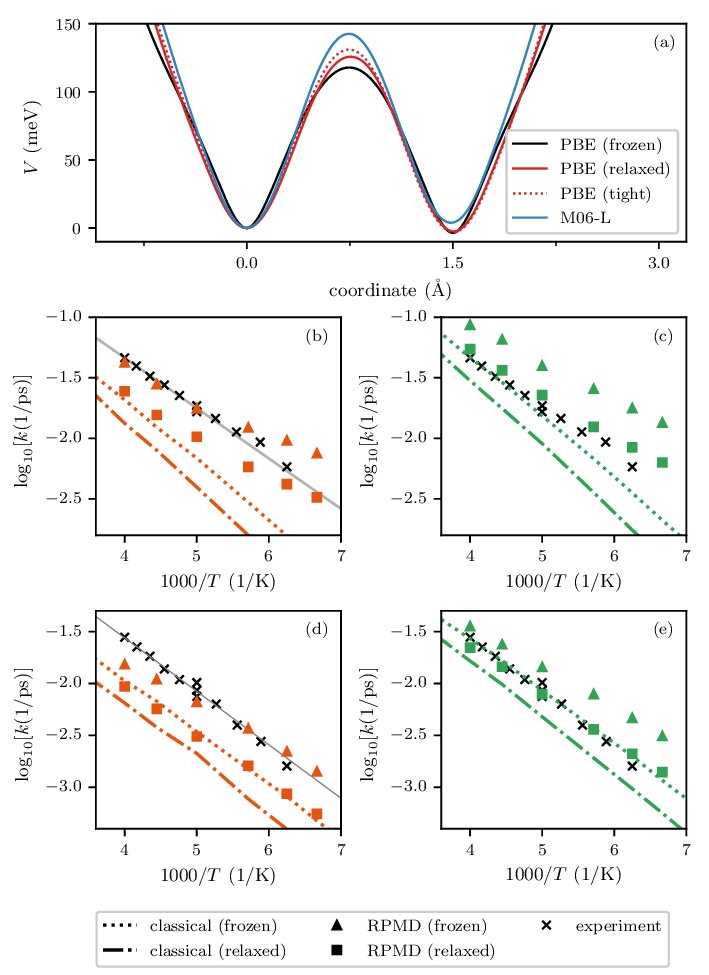}%
	\caption{(a) Double-well potentials along the reaction coordinate calculated using the PBE exchange--correlation functional with \emph{modified light} settings for a 6-layer metal slab frozen at the bare-slab geometry (black) and with the top three layers allowed to relax (red). The dotted red line shows the potential energy recomputed along the relaxed reaction path using PBE with  \emph{default tight} settings. Also shown are the results for relaxed geometries and energies computed using the M06-L exchange--correlation functional~\cite{zhaoNewLocal2006} with \emph{default tight} settings.
	(b,c) Hydrogen escape rates from the \emph{hcp} well for an electronic friction spectrum broadened by \SI{0.02}{eV} and for an Ohmic spectrum, respectively. Potentials computed using the PBE functional for a frozen and a relaxed metal substrate were used as specified in the legend. The grey line shows the linear fit to the experimental rates measured between 200 and \SI{250}{K}. (d,e) Same for deuterium.
	\label{fig:pes-conv}}
\end{figure}

The model PESs were constructed as described in the caption to \fig{pes-build}. 
To assess the effect of surface relaxation, we optimized the geometry of the top three metal layers at every point along the reaction coordinate and recomputed the escape rates using an updated model PES, marked ``PBE (relaxed)'' in \fig{pes-conv}(a). The relaxed model was constructed using the modified light settings for the DFT calculations. Using the default tight results in comparatively small changes, shown with a dotted line in 
\fig{pes-conv}(a). The relaxed model still identifies the \emph{fcc} hollow as the more stable site but predicts a reaction barrier that is higher by \SI{7}{meV}, lower reactant/product well frequencies and higher barrier frequency. Combined, these factors lead to the modified rates shown with dash-dotted lines and square markers in \fig{pes-conv}(b--e) (also \emph{cf.}~Fig.~3(c,d) in the main text).

We also relaxed the surface geometry along the reaction coordinate using the \mbox{M06-L} meta-GGA exchange--correlation functional~\cite{zhaoNewLocal2006}, which is in good agreement with the experimentally measured reaction barrier for the dissociative chemisorption of \ce{H2} on Cu(111)~\cite{oudotReactionBarriers2024}. During relaxation, we used the same basis set, integration grid, $k$-grid, and dispersion corrections as for PBE. We then recomputed the potential energies using a 12$\times$12$\times$1 \mbox{$k$-grid} and the \emph{default tight} settings, resulting in the blue model potential in \fig{pes-conv}(a). Compared to the relaxed PBE model, the reaction barrier increases by \SI{17}{meV} and the energy difference between the \emph{hcp} and \emph{fcc} sites changes from \SI{-2.5}{meV} to \SI{+3.9}{meV}. These variations are similar in magnitude to the differences between the frozen and relaxed PBE models, hence, their effect on rates would be of a similar magnitude as that seen in \fig{pes-conv}(b--e).

\section{Ring-polymer generalized Langevin dynamics}

We begin with the classical GLE corresponding to the system-bath potential in Eq.~(5)~\cite{weissQuantumDissipative2012,tuckermanStatisticalMechanics2010,nitzanChemicalDynamics2006},
\begin{equation}
    \label{eq:gle-cl}
    \dv{P(t)}{t} = -\pdv{\Vext(Q)}{Q} - \!\int_{-\infty}^{t} \!\! \eta(Q, t, t') \, \frac{P(t')}{m} \dd{t'}
    +  \pdv{\coupling(Q)}{Q} \zeta(t),
\end{equation}
where we have introduced the memory--friction kernel
\begin{equation}
	\label{eq:kernel}
	\kernel(Q, t, t') = \pdv{\coupling(t)}{Q} \resolvent(t-t')  \pdv{\coupling(t')}{Q},
\end{equation}
and the resolvent
\begin{equation}
	\label{eq:resolvent}
	\resolvent(t-t') = \frac{2}{\pi} \int_{0}^{\infty} \Lambda(\omega) \cos(\omega t) \dd{\omega},
\end{equation}
related to the coloured noise $\zeta(t)$ by the second fluctuation--dissipation theorem \cite{kuboFluctuationdissipationTheorem1966}
\begin{equation}
	\expval{\zeta(t) \zeta(t')} = \kB T K(t-t').
\end{equation}
For convenience, in \eqn{kernel} we adopt the shorthand notation
\begin{equation}
    \pdv{\coupling(t)}{Q} \equiv \pdv{\coupling(Q(t))}{Q(t)}. 
\end{equation}
Next, we define the normal-mode coordinates that diagonalize the spring term in Eq.~(3),
\begin{equation}
    \label{eq:nm-def}
	\xnm[n]_{\nu} = 
		N^{-1/2} \sum_{l} T_{ln} q^{(l)}_{\nu} 
		\quad \Leftrightarrow \quad
	q^{(l)}_{\nu} = 
		N^{1/2} \sum_{n} T_{ln} \xnm[n]_{\nu}
\end{equation}
where
\begin{equation}
	\label{eq:nm-trans}
	T_{ln} = \begin{cases}
		N^{-1/2} & n = 0 \\
		(2/N)^{1/2} \sin(\pi l n / N) & n = 1, \dotsc, \lambda_{N} \\
		(2/N)^{1/2} \cos(\pi l n / N) & n = -\lambda_{N}, \dotsc, -1 \\
		(-1)^{l} N^{-1/2} & n = -N/2
	\end{cases}
\end{equation}
with $\lambda_{N} = \floor[(N-1)/2]$ and $n = 0,\, \pm 1,\, \pm 2,\, \dotsc \; $. 
The prefactor $N^{-1/2}$ is included in \eqn{nm-def} so that the normal-mode coordinate
$\vb{q}^{(0)}$ corresponds to the ring-polymer centre of mass (centroid). In the transformed coordinates,
the spring potential reads
\begin{align}
    \nmS(\vbxnm_{\nu}) & = \frac{1}{N} S_N(\vb{q}_{\nu}) = \sum_{n} \frac{m_{\nu} \wnm[n]^{2}}{2} \qty(\xnm[n]_{\nu})^2,
\end{align}
with the normal-mode frequencies $\wnm[n]$ given in Eq.~(9). We also define a normal-mode interaction potential
\begin{equation}
    \label{eq:nmF-def}
	\nm{F}[n] = 
		N^{-1/2} \sum_{l} T_{ln} F(Q^{(l)})
		\quad \Leftrightarrow \quad
	F(Q^{(l)}) = 
		N^{1/2} \sum_{n} T_{ln} \nm{F}[n]
\end{equation}
which allows us to cast the ring-polymerized version of Eq.~(5) as
\begin{subequations}
	\begin{align}
		\frac{1}{N} \left\{
			 \sum_{l} V(\vb{q}^{(l)}) + S_N(\vb{Q})
			 + \sum_{\nu=1}^{\mathclap{\nbath}} S_N(\vb{q}_{\nu})
			 \right\} & = \nmVsys(\nm{\vb{Q}}) +  \sum_n \sum_{\nu=1}^{\mathclap{\nbath}} 
			\left\{
					\frac{m \omega_{\nu}^{2}}{2}\qty[
						\nm{q}[n]_{\nu} - \frac{c_{\nu} \nm{F}[n]}{m \omega_{\nu}^{2}}
					]^{\mathrlap{2}}
				+ \frac{m \wnm[n]^2}{2} \qty[\nm{q}[n]_{\nu}]^2
			\right\} 
		\\
		\nmVsys(\nm{\vb{Q}}) & = \nmVext(\nm{\vb{Q}}) + \sum_{n} \frac{m \wnm[n]^2}{2} \qty[\nm{Q}[n]]^2 \\
		\nmVext(\nm{\vb{Q}}) & = \frac{1}{N} \sum_{l} \Vext \!\qty(
			N^{1/2} \sum_{n} T_{ln} \nm{Q}[n]).
	\end{align}		
\end{subequations}
The dynamics of the bath mode $\nm{q}[n]_{\nu}$ are those of a driven harmonic oscillator,
\begin{equation}
	m\dv[2]{\nm{q}[n]_{\nu}}{t} + m \wnm[n,\nu]^{2} \nm{q}[n]_{\nu} = c_{\nu} \nm{F}[n],
\end{equation}
with frequency $\wnm[n,\nu]^{2} = \wnm[n]^{2} + \omega_{\nu}^{2}$, which can be solved
for initial conditions at $t = 0$ using standard techniques~\cite{tuckermanStatisticalMechanics2010},
\begin{equation}
	\begin{multlined}
		\nm{q}[n]_{\nu}(t) = \frac{c_{\nu}}{m \wnm[n,\nu]^{2}} \nm{F}[n](t) + 
		\cos(\wnm[n,\nu] t) \qty[
			\nm{q}[n]_{\nu}(0) - \frac{c_{\nu}}{m \wnm[n,\nu]^{2}} \nm{F}[n](0)
		] +  \sin(\wnm[n,\nu] t) \frac{\nm{p}[n]_{\nu}(0)}{m \wnm[n,\nu]} \\
		- \frac{c_{\nu}}{m \wnm[n,\nu]^{2}} \int_{0}^{t} 
		\cos[\wnm[n,\nu] (t-t')] \sum_{n''} \pdv{\nm{F}[n](t')}{\nm{Q}[n'']} \frac{\nm{P}[n''](t')}{m} \dd{t'}.
	\end{multlined}
\end{equation}
Substituting this result into the equation of motion for the system normal-mode momentum $\nm{P}[n']$ gives
\begin{align}
	\dv{\nm{P}[n']}{t} & = -\pdv{\nmVsys}{\nm{Q}[n']} + \sum_{n,\nu} c_{\nu} \pdv{\nm{F}[n]}{\nm{Q}[n']} \qty[
		\nm{q}[n]_{\nu} - \frac{c_{\nu} \nm{F}[n]}{m \omega_{\nu}^{2}}
	]  \\
	& = -\pdv{\nmVsys}{\nm{Q}[n']} - \sum_{n} \qty(\nm{F}[n] \pdv{\nm{F}[n]}{\nm{Q}[n']} \sum_{\nu} \frac{c_{\nu}^{2}}{m \omega^{2}_{\nu}} 
		\frac{\wnm[n]^2}{\omega^{2}_{\nu} + \wnm[n]^2}
	) \nonumber \\
	\label{eq:rpmd-gle-pos}
	& \phantom{= {} }- \sum_{n, n''} \pdv{\nm{F}[n](t)}{\nm{Q}[n']} \int_{0}^{t} 
	\qty(
		\sum_{\nu} \frac{c_{\nu}^{2}}{m \wnm[n,\nu]^{2}} 
		\cos[\wnm[n,\nu] (t-t')] ) 
	\pdv{\nm{F}[n](t')}{\nm{Q}[n'']} \frac{\nm{P}[n''](t')}{m} \dd{t'} \\
	& \phantom{= {} }+ \sum_{n,\nu} c_{\nu} \pdv{\nm{F}[n](t)}{\nm{Q}[n']} 
	\qty{
		\cos(\wnm[n,\nu] t) \qty[
			\nm{q}[n]_{\nu}(0) - \frac{c_{\nu}}{m \wnm[n,\nu]^{2}} \nm{F}[n](0)
		] +  \sin(\wnm[n,\nu] t) \frac{\nm{p}[n]_{\nu}(0)}{m \wnm[n,\nu]} 
	}. \nonumber
\end{align}
We identify the first line of \eqn{rpmd-gle-pos} with the force corresponding to the renormalized ring-polymer potential,
\begin{equation}
	\nmVnorm(\nm{\vb{Q}}) = \nmVsys(\nm{\vb{Q}}) + \frac{1}{2} \sum_{\smash[b]{n}} \alpha^{(n)} \qty\big[\nm{F}^{(n)}]^2,
	\label{eq:rp-ren-pot}
\end{equation}
where $\smash{\alpha^{(n)}}$ is defined in Eq.~(8b) of the main text. The new forces arising from coupling to the bath, 
\begin{equation}
	-\pdv{\nm{Q}[n']} \qty(
		\nmVnorm - \nmVsys
	) =  -\sum_{n} \alpha^{(n)} \nm{F}[n] \qty{ 
        \sum_{l} T_{ln} T_{ln'} \pdv{F(Q^{(l)})}{Q^{(l)}} },
\end{equation}
can be calculated efficiently by pre-computing $\alpha^{(n)}$ and $T_{ln} T_{ln'}$. For a linear interaction, only the diagonal ($n = n'$) terms survive. The second line of \eqn{rpmd-gle-pos} defines the frictional forces, with the sum over bath modes (in parentheses) equal to the resolvent in Eq.~(12). The last line defines the stochastic forces
\begin{equation}
	\sum_n \pdv{\nm{F}[n](t)}{\nm{Q}[n']} \zeta_n(t)
\end{equation}
that can be shown~\cite{Lawrence2019a} to satisfy
\begin{equation}
	\label{eq:rpmd-fd2}
	\expval{\zeta_n(t) \zeta_{n'}(t')} = \kB T \delta_{n,n'} \delta(t-t').
\end{equation}
To complete the derivation, we shift the time origin from $t = 0$ to $t = -\infty$, so that \eqn{rpmd-gle-pos} agrees with \eqn{gle-cl} in the classical limit ($N = 1$).

\section{Auxiliary variable propagation%
\label{sec:auxvar-propa}}

\subsection{Propagation algorithm%
\label{sec:auxvar-alg}}

In what follows, we use the mass-weighted coordinates $ \uQ = \sqrt{m} Q $ and 
$ \uP = P / \sqrt{m} $. The corresponding RPMD GLE reads
\begin{equation}
	\label{eq:mw-rpmd-gle}
	\dv{\nm{\uP}[n']}{t} = -\pdv{\nmVnorm(t)}{\nm{\uQ}[n']} - 
	\sum_{n,n''} \pdv{\nm{F}[n](t)}{\nm{\uQ}[n']}\!\int_{-\infty}^{t} \!\! K^{(n)}(t-t') \, \pdv{\nm{F}[n](t')}{\nm{\uQ}[n'']} \nm{\uP}[n''](t') \dd{t'}
    + \sum_{n} \pdv{\nm{F}[n](t)}{\nm{\uQ}[n']} \zeta_n(t),
\end{equation}
which we want to replace by a system of Markovian stochastic differential equations
\begin{subequations}
    \begin{align}
        \dv{\nm{\uP}[n']}{t} & = -\pdv{\nmVnorm(t)}{\nm{\uQ}[n']}
		- \sum_{n} \pdv{\nm{F}[n](t)}{\nm{\uQ}[n']} \nm{\bm{\theta}}[n]{}^{\intercal} \,
		\nm{\vb{s}}[n]
        \label{eq:mw-OU-P} \\
        \dv{\nm{\vb{s}}[n']}{t} & = 
			-\nm{\vb{A}}[n'] \nm{\vb{s}}[n'](t) 
			+\nm{\vb{B}}[n'] \nm{\bm{\xi}}[n'](t)  
			+\nm{\vb{f}}[n'](t)
		\label{eq:mw-OU-s}
    \end{align}
\end{subequations}
where $\nm{\vb{A}}[n']$ and $\nm{\vb{B}}[n']$ are constant matrices,
$\nm{\bm{\theta}}[n]$ are constant vectors,  $\nm{\bm{\xi}}[n']$ are vectors whose entries are Gaussian random variates with zero mean and unit variance, $\nm{\vb{f}}[n']$ are as yet unknown driving forces, and $\nm{\vb{s}}[n']$ are the auxiliary variables. 
Assuming the real parts of the eigenvalues of $\nm{\vb{A}}[n']$ are all positive, \eqn{mw-OU-s} with initial conditions in the infinite past are solved by~\cite{gardinerHandbookStochastic1985}
\begin{equation}
    \nm{\vb{s}}[n'] = \int_{-\infty}^{t} \eu[-\nm{\vb{A}}[n'](t-t')] \qty(
			\nm{\vb{B}}[n'] \nm{\bm{\xi}}[n'](t') + \nm{\vb{f}}[n'](t')
		) \dd{t'}.
\end{equation}
Substituting this into \eqn{mw-OU-P} gives
\begin{equation}
\begin{multlined}
	\dv{\nm{\uP}[n']}{t} = -\pdv{\nmVnorm(t)}{\nm{\uQ}[n']}
		- \sum_{n} \pdv{\nm{F}[n](t)}{\nm{\uQ}[n']} \int_{-\infty}^{t} 
		\nm{\bm{\theta}}[n]{}^{\intercal} \,
		\eu[-\nm{\vb{A}}[n](t-t')]  \nm{\vb{f}}[n](t') \dd{t'} \\
		- \sum_{n} \pdv{\nm{F}[n](t)}{\nm{\uQ}[n']} \int_{-\infty}^{t} 
		\nm{\bm{\theta}}[n]{}^{\intercal} \,
		\eu[-\nm{\vb{A}}[n](t-t')] \nm{\vb{B}}[n] \nm{\bm{\xi}}[n](t') \dd{t'}.
    \label{eq:OU-subbed}
\end{multlined}	
\end{equation}
Comparing \eqs{mw-rpmd-gle}{OU-subbed} we conclude that
\begin{equation}
    \nm{\vb{f}}[n](t') = \sum_{n''} \nm{\bm{\vartheta}}[n] \, \pdv{\nm{F}[n](t')}{\nm{\uQ}[n'']} 
	\nm{\uP}[n''](t'),
\end{equation}
where we introduce another constant vector $\nm{\bm{\vartheta}}[n]$, so that
\begin{equation}
    K^{(n)}(t-t') =
	 \nm{\bm{\theta}}[n]{}^{\intercal} \,
	 \eu[-\nm{\vb{A}}[n](t-t')] 
	 \nm{\bm{\vartheta}}[n].
\end{equation}
We also identify
\begin{equation}
    \zeta_n(t) = -\int_{-\infty}^{t} \nm{\bm{\theta}}[n]{}^{\intercal} \,
	\eu[-\nm{\vb{A}}[n](t-t')] \nm{\vb{B}}[n] \nm{\bm{\xi}}[n](t') \dd{t'},
\end{equation}
which must satisfy the fluctuation--dissipation theorem in \eqn{rpmd-fd2}. This amounts to imposing
\begin{equation}
    \int_{-\infty}^{t+\tau} \!\! \dd{t''} 
	\int_{-\infty}^{t} \!\!\! \dd{t'} 
	\nm{\bm{\theta}}[n]{}^{\intercal} \, \eu[-\nm{\vb{A}}[n](t+\tau-t'')] \nm{\vb{B}}[n] \expval{
		\nm{\bm{\xi}}[n](t'') \nm{\bm{\xi}}[n](t')^{\intercal}
	} \nm{\vb{B}}[n]{}^{\intercal} \, \eu[-\nm{\vb{A}}[n]{}^{\intercal}(t-t')]
	\nm{\bm{\theta}}[n] = \kB T \nm{\bm{\theta}}[n]{}^{\intercal} \,
	\eu[-\nm{\vb{A}}[n]\tau] 
	\nm{\bm{\vartheta}}[n].
\end{equation}
Given that $ \expval{\nm{\bm{\xi}}[n](t'') \nm{\bm{\xi}}[n'](t')^{\intercal}} = \delta_{n,n'} \delta(t''-t') \vb{I} $, where $\vb{I}$ is the identity matrix, this reduces to
\begin{equation}
    \int_{-\infty}^{t} \!\! \dd{t'} 
	\nm{\bm{\theta}}[n]{}^{\intercal} \, \eu[-\nm{\vb{A}}[n](t+\tau-t')] \nm{\vb{B}}[n] \nm{\vb{B}}[n]{}^{\intercal} \, \eu[-\nm{\vb{A}}[n]{}^{\intercal}(t-t')]
	\nm{\bm{\theta}}[n] = \kB T \nm{\bm{\theta}}[n]{}^{\intercal} \,
	\eu[-\nm{\vb{A}}[n]\tau] 
	\nm{\bm{\vartheta}}[n].
\end{equation}
Since
\begin{equation}
    \dv{t'} \qty(
        \eu[-\vb{A}(t+\tau-t')] \vb{C} \eu[-\vb{A}^{\intercal} (t-t')]
    ) = \eu[-\vb{A}(t+\tau-t')] \qty(
        \vb{A} \vb{C} + \vb{C} \vb{A}^{\intercal}
    ) \eu[-\vb{A}^{\intercal} (t-t')]
\end{equation}
if we choose $ \nm{\vb{A}}[n] \nm{\vb{C}}[n] + \nm{\vb{C}}[n] \nm{\vb{A}}[n]{}^{\intercal}  =  \nm{\vb{B}}[n]\nm{\vb{B}}[n]{}^{\intercal} / \kB T $, the condition takes the form
\begin{equation}
    \nm{\bm{\theta}}[n]{}^{\intercal} \, \eu[-\nm{\vb{A}}[n]\tau] \nm{\vb{C}}[n] 
	\nm{\bm{\theta}}[n]
     = \nm{\bm{\theta}}[n]{}^{\intercal} \, \eu[-\nm{\vb{A}}[n]\tau] 
	 \nm{\bm{\vartheta}}[n],
\end{equation}
satisfied for
\begin{equation}
    \nm{\bm{\vartheta}}[n] = \nm{\vb{C}}[n] \nm{\bm{\theta}}[n],
\end{equation}
where $\nm{\vb{C}}[n] $ are real, symmetric, positive-definite matrices that are proportional to the covariance of the auxiliary variables in the absence of coupling to the system. Without loss of generality, we may set all these matrices equal to the identity~\cite{ceriottiPhD2010}, so that the parameters for propagating the target GLE
can be found by fitting 
\begin{equation}
    K^{(n)}(\tau) = \nm{\bm{\theta}}[n]{}^{\intercal} \, \eu[-\nm{\vb{A}}[n]\tau] 
	\nm{\bm{\theta}}[n],
\end{equation}
where the left-hand side is given by Eq.~(12), and $(\nm{\bm{\theta}}[n]$, 
$\nm{\vb{A}}[n])$ are parameters to be varied, subject to the constraint that 
$\nm{\vb{A}}[n]$ have a spectrum with a positive real part. The final set of equations of motion is then
\begin{subequations}
    \begin{align}
		\dv{\nm{\uP}[n']}{t} & = -\pdv{\nmVnorm(t)}{\nm{\uQ}[n']}
		- \sum_{n} \pdv{\nm{F}[n](t)}{\nm{\uQ}[n']} \nm{\bm{\theta}}[n]{}^{\intercal} \,
		\nm{\vb{s}}[n]
		\label{eq:aux-eom-P} \\
		\dv{\nm{\vb{s}}[n']}{t} & = 
				-\nm{\vb{A}}[n'] \nm{\vb{s}}[n'](t) 
				+\nm{\vb{B}}[n'] \nm{\bm{\xi}}[n'](t)  
				+ \sum_{n''} \pdv{\nm{F}[n'](t')}{\nm{\uQ}[n'']} \,
				\nm{\bm{\theta}}[n']  \nm{\uP}[n''](t').
		\label{eq:aux-eom-s}
	\end{align}
\end{subequations}
These are propagated according to an algorithm that we based on \refs{Ceriotti2010}{stellaGeneralizedLangevin2014}. A single propagation step evolving the system through a time increment $\tau$ is notated as
\begin{equation}
	\mathcal{O}_s^{(\tau/2)} 
	\mathcal{B}_{P,V}^{(\tau/2)}
	\mathcal{B}_{P,F}^{(\tau/2)}
	\mathcal{A}_{\omega}^{(\tau)}
	\mathcal{B}_{s}^{(\tau)}
	\mathcal{B}_{P,F}^{(\tau/2)}
	\mathcal{B}_{P,V}^{(\tau/2)}
	\mathcal{O}_s^{(\tau/2)},
\end{equation}
where the calligraphic letters denote coordinate updates, to be executed in order from left to right. The updates are
\begin{align}
	& \mathcal{O}_s^{(\tau/2)} : & 
	\nm{\vb{s}}[n'] & \leftarrow \nm{\vb{T}}[n']_{\tau/2} \,  \nm{\vb{s}}[n'] + \nm{\vb{S}}[n']_{\tau/2} \, \nm{\bm\xi}[n'],  &\\
	& \mathcal{B}_{P,V}^{(\tau/2)} : & 
	\nm{P}[n']  & \leftarrow \nm{P}[n'] - \frac{\tau}{2} \pdv{\nm{Q}[n']} \qty(
		\nmVext(\nm{\vb{Q}}) + \frac{1}{2} \sum_n \alpha^{(n)} [\nm{F}^{(n)}]^{2}
	), & \\
	& \mathcal{B}_{P,F}^{(\tau/2)} : & 
	\nm{\uP}[n']  & \leftarrow \nm{\uP}[n'] \!\! - \frac{\tau}{2} \sum_{n} \pdv{\nm{F}[n]}{\nm{\uQ}[n']} \nm{\bm{\theta}}[n]{}^{\intercal} \,
	\nm{\vb{s}}[n], & \\
	& \mathcal{A}_{\omega}^{(\tau)} : & 
	\mqty(
		\nm{\uP}[n'] \\
		\nm{\uQ}[n'] 
	) & \leftarrow
	\mqty(
		\vphantom{\nm{\uP}[n']}
		\cos(\wnm[n']\tau) & -\wnm[n'] \sin(\wnm[n']\tau) \\
		\vphantom{\nm{\uQ}[n']}
		\wnm[n']^{-1} \sin(\wnm[n']\tau) & \cos(\wnm[n']\tau)
	)
	\mqty(
		\nm{\uP}[n'] \\
		\nm{\uQ}[n'] 
	), & \\
	& \mathcal{B}_{s}^{(\tau)} : & 
	\nm{s}[n']  & \leftarrow \nm{s}[n'] \! + \tau \sum_{n''} 
	\pdv{\nm{F}[n']}{\nm{\uQ}[n'']} \, \nm{\bm{\theta}}[n'] \nm{\uP}[n''], & 
\end{align}
with constant coefficient matrices
\begin{subequations}
	\label{eq:gle-propa-mats}
	\begin{gather}
		\nm{\vb{T}}[n](\tau/2) = \eu[-\nm{\vb{A}}[n]\tau/2],  \\
		\nm{\vb{S}}[n]_{\tau/2} \widetilde{\vb{S}}^{(n)\,\intercal}_{\tau/2} = 
			\kB T \qty(
				\vb{I} - 
				\nm{\vb{T}}[n]_{\tau/2} \widetilde{\vb{T}}^{(n)\,\intercal}_{\tau/2}
			)
	\end{gather}
\end{subequations}
Following \refx{stellaGeneralizedLangevin2014}, we restrict the matrices $\nm{\vb{A}}[n]$ to be (2$\times$2)-block-diagonal, with the $k$-th block taking the form
\begin{equation}
    \qty\big[\nm{\vb{A}}[n]]_k = \mqty[
        1/\tau_k^{(n)} & -\omega_k^{(n)} \\
        \omega_k^{(n)} & 1/\tau_k^{(n)}
    ],
\end{equation}
so that the matrices in \eqn{gle-propa-mats} are also block diagonal,
\begin{subequations}
	\label{eq:gle-propa-mats2}
	\begin{gather}
		\qty\big[\nm{\vb{T}}[n]]_k(\tau/2) = 
			\exp(-\tau/2 \tau_k^{(n)}) \mqty[
				\cos(\tfrac{1}{2} \tau \omega_k^{(n)}) & \sin(\tfrac{1}{2} \tau \omega_k^{(n)}) \\
				-\sin(\tfrac{1}{2} \tau \omega_k^{(n)}) & \cos(\tfrac{1}{2} \tau \omega_k^{(n)})
			  ]  \\
		\qty\big[\nm{\vb{S}}[n]]_k(\tau/2) = 
		\vb{I} \sqrt{{\kB T \qty(1 - \exp(-\tau/\tau_k^{(n)}))}}.
	\end{gather}
\end{subequations}
By the same token, the coefficient vectors are restricted to the form $
	\nm{\bm{\theta}}[n]_{2k-1} = c^{(n)}_{k}\!\!,\ $
	$\nm{\bm{\theta}}[n]_{2k} = 0$, with $k = 1,\,\ldots,\,\naux$.

\subsection{Parameter fitting%
\label{sec:auxvar-fit}}

Given the restrictions imposed on $\nm{\vb{A}}[n]$ and $\nm{\bm{\theta}}[n]$, the fitting equation becomes
\begin{equation}
	K^{(n)}(\tau) = \sum_{k = 1}^{\naux} \qty[c^{(n)}_{k}]^{2} \exp(-\tau/\tau_k^{(n)}) \cos(\tau \omega_k^{(n)})
\end{equation}
We find suitable values of $c^{(n)}_{k}$, $\tau_k^{(n)}$, and $\omega_k^{(n)}$ by calculating the resolvent on a uniform grid ranging from $\tau = 0$ to $\tau = \tau_{\mathrm{max}}$ and performing nonlinear least-squares minimization on the residuals
\begin{equation}
	\rho_i = 
	K^{(n)}(\tau_i) - \sum_{k = 1}^{\naux} d_{k} \exp(- \qty[\sum_{\nu=1}^{k} \gamma_{\nu} ] \tau_i) \cos(\tau_{i} \omega_k^{(n)}),
\end{equation}
where the form of the objective function was chosen to eliminate trivial degenerate solutions and to have a simple analytical expression for the Jacobian.

\subsubsection{Exponentially-dampled Ohmic spectral density}

\begin{figure}[h]
	\includegraphics{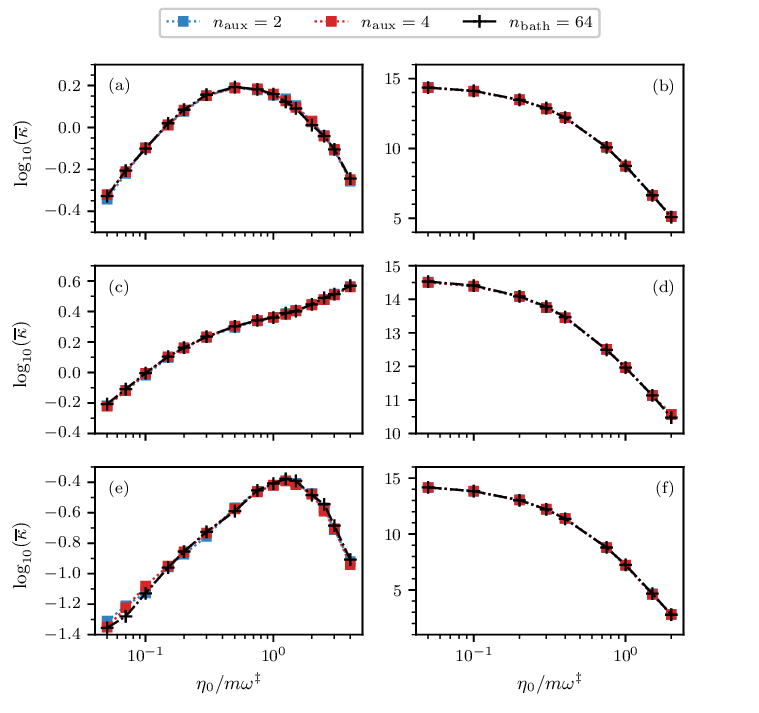}%
	\caption{RPMD rates for the double-well model  from \refx{bridgeQuantumRates2024} with uniform \mbox{(a--b)}, symmetric \mbox{(c--d)}, and asymmetric \mbox{(e--f)} position-dependent exponentially damped Ohmic friction ($\wwc = \SI{500}{\wn}$). We compare the rates calculated using a harmonic bath representation of the friction with $n_\text{bath} = 64$ modes to those computed using the analytical renormalized ring-polymer potential [\eqn{rp-ren-pot}] with an auxiliary-variable propagator for the RPMD GLE [\eqn{mw-rpmd-gle}]. Panels (a), (c), and (d) are for $T = \SI{300}{K}$, $N = 16$, whereas (b), (d), and (f) are for $T = \SI{50}{K}$, $N = 64$.
	\label{fig:expohmic-naux}}
\end{figure}

For the exponentially-damped Ohmic spectral density
\begin{equation}
	J(\omega) = \eta_0 \omega \eu[-\omega/\omega_c],
\end{equation}
we can define a reduced resolvent
\begin{equation}
	\resolvent[n]_{*}(t) = \frac{2}{\pi}\! \int_{\wnm[n]^{*}}^{\infty} \!
	\frac{f_n(u)}{u} \, \eu[-f_n(u)] \cos(u t^{*}) \dd{u},
\end{equation}
where $\wnm[n]^{*} = \wnm[n]/\wwc$ and $ t^{*} = \wwc t$ are the reduced $n$-th ring-polymer normal-mode frequency and reduced time, respectively, and
\begin{equation}
	f_n(u) = \sqrt{u^2 - [\wnm[n]^{*}]^2}.
\end{equation}
We calculated the reduced resolvents for normal-mode frequencies $0 \leq \wnm[n]^{*} \! < 25$ and fitted propagator coefficients for $\naux = 1,\,2,\,3,\,4$, storing the results in a database. Given a set of ring-polymer normal-mode frequencies corresponding to temperature $T$ and bead number $N$, we compute the corresponding coefficients from a cubic spline interpolation of the stored data. Conversion to system units is done according to the relations
\begin{align}
	c_{k}^{(n)} & = c_{k,*}^{(n)} \sqrt{\eta_{0} \omega_{c}}, &
	\omega_{k}^{(n)}  & = \omega_{k,*}^{(n)} \wwc, &
	\tau_{k}^{(n)} & = \tau_{k,*}^{(n)} / \wwc.
\end{align}
Using the double-well model and the three position-dependent friction profiles in \refx{bridgeQuantumRates2024}, we performed extensive testing of our propagation algorithm at a range of temperatures, bead numbers and friction strengths, finding essentially perfect agreement for $\naux \geq 2$. \Fig{expohmic-naux} shows the agreement between RPMD rates calculated using different bath representations for $T = \SI{300}{K}$ and $\SI{50}{K}$ [see Eq.~(25) of \refx{bridgeQuantumRates2024} for the definition of $\overline{\kappa}$].

\subsubsection{Ab~initio spectral density}

We found that the spatial variation of the electronic friction tensor has a negligible effect on the rates computed for H/D diffusing on Cu(111). Therefore, all calculations we quote for this system were performed assuming a spatially uniform friction ($\nm{F}[n](\nm{\vb{Q}}) = \nm{Q}[n]$).
To simplify the harmonic discretization~\cite{waltersDirectDetermination2017} and the fitting of auxiliary variable parameters, we applied a broad exponential damping window to the \emph{ab~initio} friction spectrum at the \emph{fcc} site [dotted line in Fig.~1(c)]. We fitted coefficients to the reduced resolvents
\begin{equation}
	\resolvent[n]_{*}(t) = \frac{2}{\pi} \int_{0}^{\infty} \!
	\eu[-\omega/\wwc] \times \frac{\Lambda(\omega)}{\Lambda(0)} \frac{ \omega^2}{\omega^2 + \wnm[n]^2} \cos(
		t \sqrt{\omega^2 + \wnm[n]^2}
	) \dd{\omega},
\end{equation}
considering damping windows (first factor in the integrand) of width $\wwc = \SI{2000}{\wn}$ and $\SI{4000}{\wn}$. The fits were separately computed for every ring-polymer normal-mode frequency appearing in our calculations. The coupling coefficients were scaled as $\smash{c_{k}^{(n)}} = \smash{c_{k,*}^{(n)} \Lambda^{1/2}(0)}$, where $\Lambda(0)$ is the target value of the static friction. No scaling had to be applied to the remaining propagation parameters. 

\begin{figure}[h!]
	\includegraphics{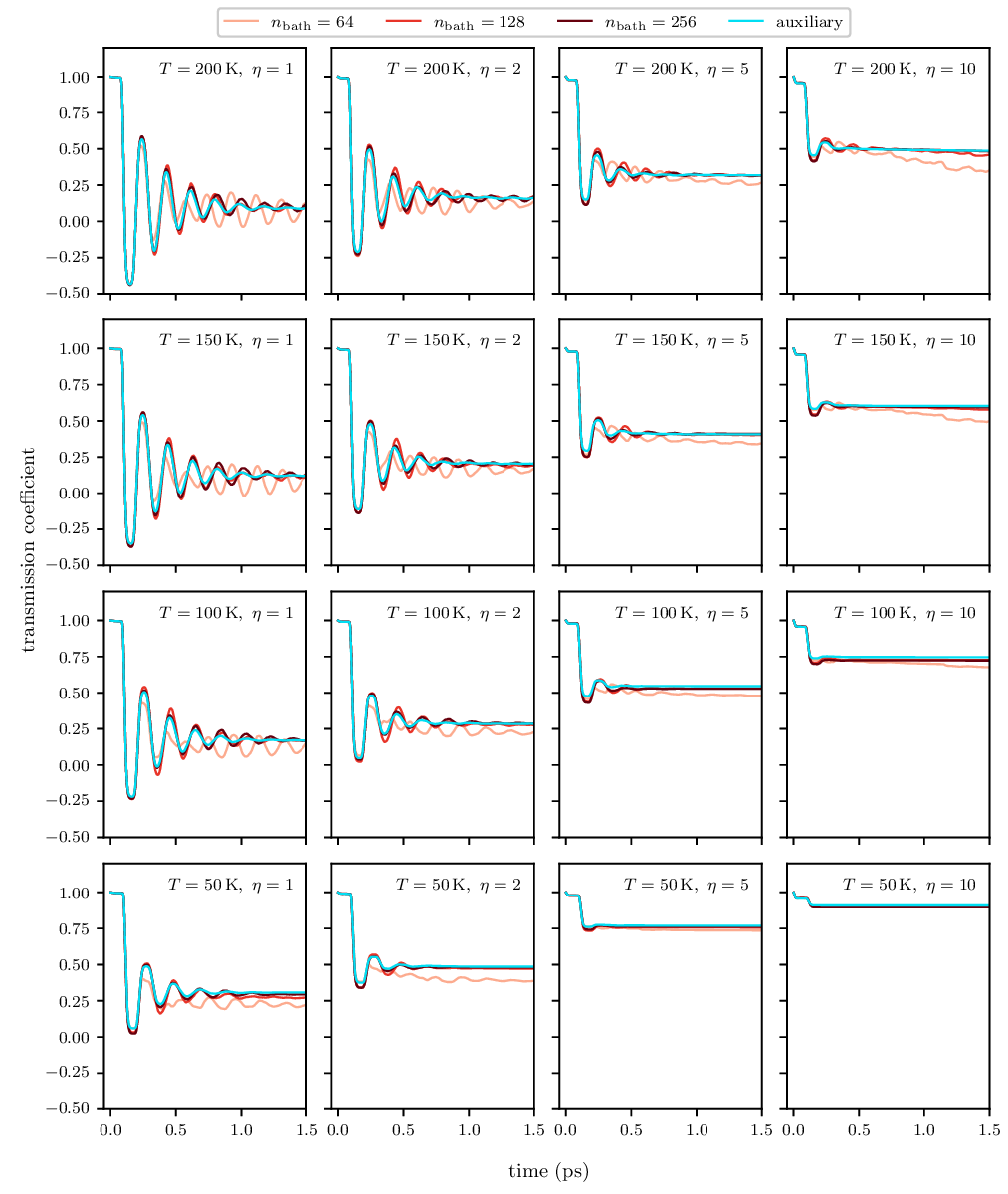}%
	\caption{Classical ($N=1$) transmission coefficients for the fcc friction spectrum in Fig.~1(c), damped by  $w(\omega) = \eu[-\omega/\omega_c]$, $\omega_c = \SI{2000}{\wn}$. The simulation temperatures and scaling factors applied to the \emph{ab~initio} friction are written on the plot panels. Calculations using harmonic bath discretization \cite{waltersDirectDetermination2017} are shown in shades of red (see legend for the number of bath modes). Results from the direct propagation of the GLE mapped onto $\naux = 3$~pairs of auxiliary variables is shown in black.
	\label{fig:abinit-aux-F04}}
\end{figure}

To test the fit quality, we compared the classical ($N = 1$) transmission coefficients computed using auxiliary variable propagation to explicit harmonic bath calculations. The results for our best parameter sets (used in production calculations) are shown in \figs{abinit-aux-F04}{abinit-aux-F05} for a damping of 
$\wwc = \SI{2000}{\wn}$ and $\SI{4000}{\wn}$, respectively.

\begin{figure}[h!]
	\includegraphics{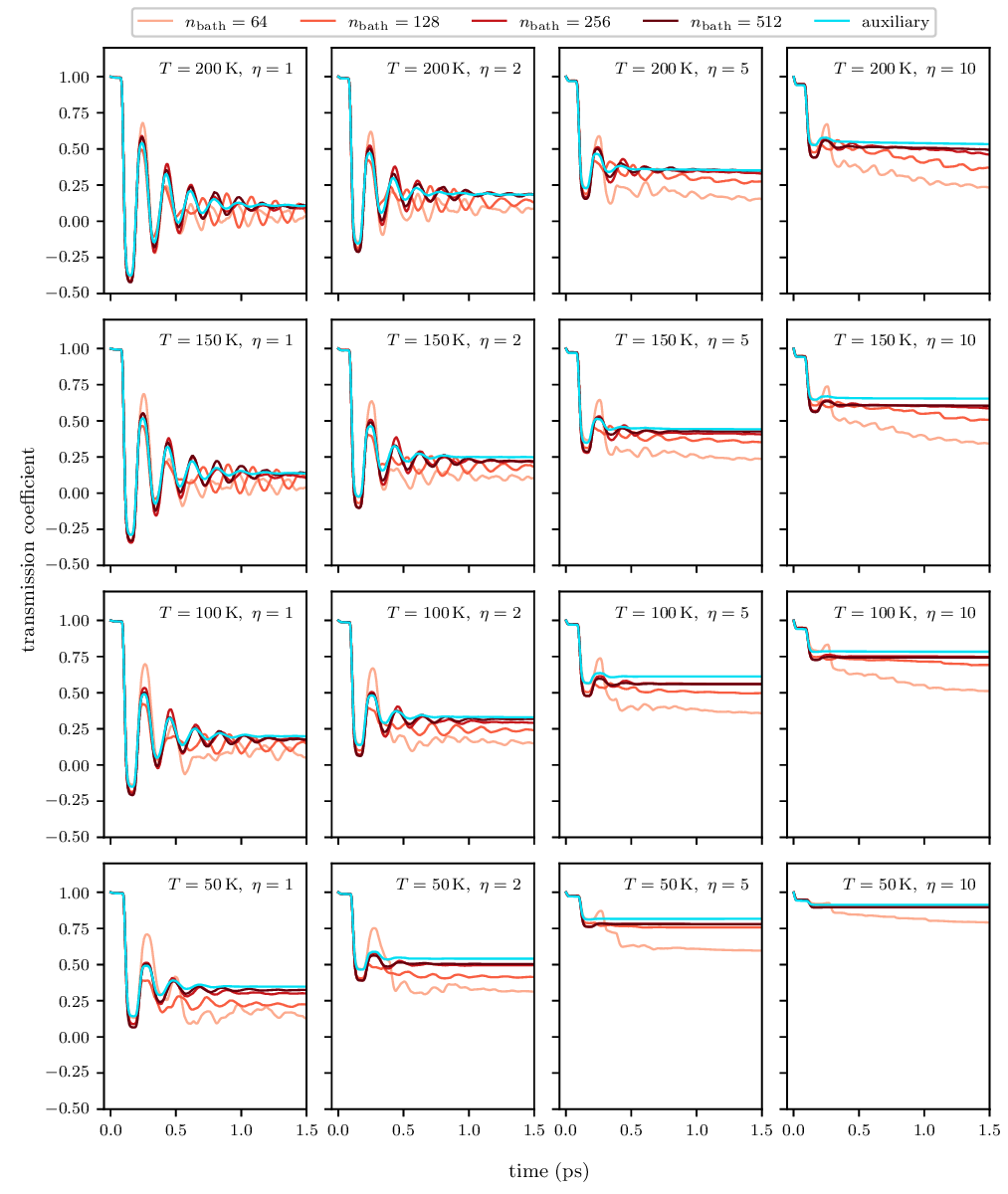}%
	\caption{Same as \fig{abinit-aux-F04} but for a cut-off window $\omega_c = \SI{4000}{\wn}$ and for $\naux = 5$.
	\label{fig:abinit-aux-F05}}
\end{figure}

\subsection{QTST calculations with the renormalized potential%
\label{sec:mf-conv}}

\begin{figure}[h]
	\includegraphics{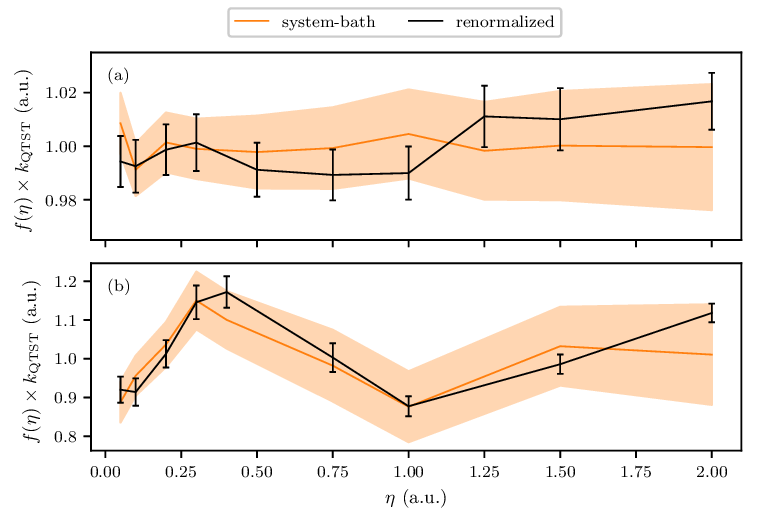}%
	\caption{QTST rates for the uniform-friction model from Ref~\cite{bridgeQuantumRates2024}, simulated with path-integral molecular dynamics (PIMD) using the harmonic system-bath representation and the mean-field (renormalized potential) representation of the dissipative environment. Since the low-temperature rates vary over many orders of magnitude with $\eta$, we multiply the calculated rates and corresponding confidence intervals by $f(\eta) = \exp[c_0 + c_1 \eta + c_2 \eta^2]$ for ease of visualization.
	(a)~QTST rates for $T = \SI{300}{K}$, $N=16$, 
	$c_0 = \num{1.694e+01}$, 
	$c_1 = \num{3.082e-02}$, and
	$c_2 = \num{5.960e-03}$. 
	(b)~QTST rates for $T = \SI{50}{K}$, $N=64$, 
	$c_0 = 32.843$, 
	$c_1 = 16.406$, and
	$c_2 = 2.655$. All values are in atomic units. In both cases, we used $n_{\text{bath}} = 64$ harmonic bath modes for the system-bath simulations. Given similar durations of thermal sampling, the confidence intervals calculated for the two representations
	at low friction and high temperature are similar. With decreasing temperature and increasing friction, the uncertainty estimates for system-bath simulation results increase due to limited sampling of the harmonic bath modes. 
	\label{fig:kqtst-conv}}
\end{figure}

In the Letter, we comment that using the renormalized potential instead of the harmonic system-bath representation of the dissipative environment substantially simplifies QTST rate calculations. The coupling coefficients in Eq.~(8b) need only be calculated once at the start of the calculation, and using the renormalized potential in Eq.~(8a) to drive the dynamics directly yields the $n_{\text{bath}} \to \infty$ limit of the dissipative QTST rate. The computational cost of a single propagation step is reduced due to the reduction in the degrees of freedom, $N\times(n_{\text{bath}} + 1)  \to N$, and sampling efficiency is  determined solely by the system dynamics. In contrast, for an explicit system-bath representation, the thermal sampling of the harmonic bath mode distribution becomes the limiting factor. This is shown in \fig{kqtst-conv}, where we plot the QTST rates for the uniform-friction model from Ref~\cite{bridgeQuantumRates2024}. At a given temperature, all else being equal, the estimated sampling uncertainties from calculations using the renormalized potential remain approximately constant, whereas the uncertainties in system-bath simulations grow with increasing friction strength.

\section{Rate calculations}

To calculate the classical and RPMD reaction rates, we followed the Bennett--Chandler approach as described in Ref.~\citenum{Collepardo-Guevara2008}. The quantum transition-state theory (QTST) rate
was expressed as
\begin{equation}
	\label{eq:kqtst-defn}
	\kQTST(T) = \frac{1}{\qty(2 \pi \beta m)^{1/2}} \Pi(Q_a) \eu[-\beta \Delta A^{\ddagger}]
\end{equation}
where $m = \SI{1.007825}{Da}$ and \SI{2.014102}{Da} for hydrogen and deuterium, respectively, and
\begin{equation}
	\label{eq:reactant-prob}
	\Pi(Q_a) = \frac{\expval*{\delta(Q_a - \nm{Q}[0])}}{\expval*{\theta(Q^{\ddagger} - \nm{Q}[0])}},
\end{equation}
with $\expval{\cdots}$ denoting a thermal average, $\theta(x)$ denoting the Heaviside step function, $Q_a$ located at the bottom of the \emph{hcp} well and $Q^{\ddagger}$ located at the top of the PES barrier. The probability density value in \eqn{reactant-prob} was computed from a kernel density estimate (KDE) using Epanechnikov kernel functions of bandwidth $w = \SI{0.005}{\AA}$ centred on the centroid positions sampled in a thermostatted path-integral molecular dynamics (PIMD) simulation. A biasing potential,
\begin{equation}
	U_{\mathrm{bias}}(\nm{Q}[0]) = \begin{cases}
		\dfrac{k}{2} \qty\big(\nm{Q}[0] - (Q^{\ddagger} - \delta))^{2} & \nm{Q}[0] > Q^{\ddagger} - \delta \\
		0 & \text{otherwise}
	\end{cases}
\end{equation}
with $\delta = \SI{0.005}{\AA}$ and $k = \SI{50}{\electronvolt \per \angstrom^2}$ was applied to the centroid coordinate, in order to restrain it to the reactant well. Trajectories were propagated on the renormalized potential $\nmVnorm$ using the PILE-L thermostat~\cite{Ceriotti2010} with centroid friction $\tau_0 = \SI{10}{fs}$ and an integration time step $\Delta t = \SI{0.5}{fs}$. For every rate calculation, 50 independent PIMD trajectories were each propagated for \SI{110}{ps}, with centroid positions sampled every \SI{25}{fs} and the first \SI{10}{ps} of every trajectory discarded.

The second factor in \eqn{kqtst-defn} depends on the free energy
\begin{equation}
	\label{eq:thermo-int}
	\Delta A^{\ddagger} = \int_{Q_a}^{Q^{\ddagger}} \expval{\pdv{\nmVnorm(\nm{\vb{Q}})}{\nm{Q}[0]}}_{\widetilde{Q}'^{(0)}} \dd{\widetilde{Q}'^{(0)}},
\end{equation}
with $\expval{\cdots}_{\widetilde{Q}'^{(0)}}$ denoting a thermal average for the centroid constrained at $\widetilde{Q}'^{(0)}$. The integral in \eqn{thermo-int} was computed by Gauss--Legendre quadrature with 10 sample points~\cite{NumRep}. We propagated 50 independent PIMD trajectories for \SI{21}{ps} at every quadrature point, using the same settings as above \emph{sans} the biasing potential. The centroid forces for computing the thermal average in \eqn{thermo-int} were sampled every \SI{25}{fs}, with the first \SI{1}{ps} of every trajectory discarded.

The dynamical transmission coefficient is given by
\begin{equation}
	\kappa(t) = \frac{
		\expval*{\delta(Q^{\ddagger} - \nm{Q}[0]) (\nm{P}[0] \! / m) \, \theta[ \nm{Q}[0](t) - Q^{\ddagger} ]}
	}{
		\expval*{\delta(Q^{\ddagger} - \nm{Q}[0]) (\nm{P}[0] \! / m) \, \theta[ \nm{P}[0]  ] }
	}
\end{equation}
and was obtained by averaging over 100~independent RPMD simulations for every rate calculation. In every independent simulation, a PILE-L--thermostatted PIMD trajectory with the centroid constrained at $Q^{\ddagger}$ was propagated using the same settings as for the QTST calculations. Every \SI{500}{fs}, the sampled ring-polymer configurations, $\nm{\vb{Q}}$, and the momenta, $\nm{\vb{P}}$, drawn from the Boltzmann distribution, were used to launch a pair of RPMD trajectories, with the initial conditions $(\pm\nm{\vb{P}},\, \nm{\vb{Q}})$. The trajectories were propagated for \SI{3}{ps} using the auxiliary variable algorithm in \sec{sec:auxvar-propa} with an integration time step of \SI{0.5}{fs}. Between 100 and 1000 pairs of independent trajectories were launched in every simulation, until the estimated relative sampling error in the averaged $\kappa(t_p)$ dropped below 3\%.

For our low-barrier, low-friction system the assumption of separation of timescales implicit in Eq.~(1) is not satisfied at high temperatures~\cite{Craig2007}. Therefore, we used the more general expression~\cite{Lawrence2019a},
\begin{equation}
	\label{eq:fancy-rate}
	k(T) = \lim_{t \to t_p} \frac{
		\kQTST^{(R)}(T) \kappa(t)
	}{
		1 - [ \kQTST^{(\text{R})}(T) + \kQTST^{(\text{P})}(T) ] \int_{0}^{t} \kappa(t') \dd{t'} 
	},
\end{equation}
where $\kQTST^{(\text{R})}(T)$ is the QTST rate for the escape from the \emph{hcp} (reactant) well, and 
$\kQTST^{(\text{P})}(T)$ is the QTST rate for the \emph{fcc} (product) well. As a function of time, the right-hand side of \eqn{fancy-rate} plateaued for all temperature and friction regimes in this study. The transmission coefficients plotted in the main article and this document are all computed as $k(T)/\kQTST^{(\text{R})}(T)$, where $k(T)$ is given by \eqn{fancy-rate}.

\subsection{Convergence with bead number}

\begin{figure}[h!]
	\includegraphics{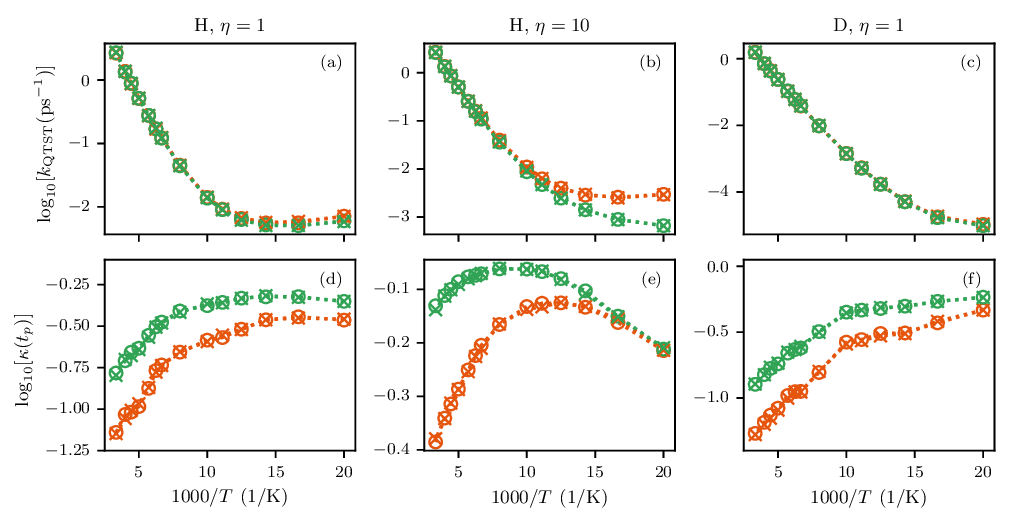}%
	\caption{(a--c) QTST rates and (d--f) transmission coefficients calculated using the smaller (open circles) and larger (crosses) bead numbers from \tab{rpmd-beads} for the same two kinds of memory--friction as in Fig.~2 of the main text. The hydrogen isotope and scaling of the friction strength are given in the headings above the top row of plot panels.
	\label{fig:bead-conv}}
\end{figure}

To assess convergence of RPMD rates with bead number, we performed rate calculations using the different numbers of beads in \tab{rpmd-beads}. It can be seen from \fig{bead-conv} that adequate convergence is achieved already for the lower number. All calculation results in the main article are quoted for the higher of the two bead numbers considered at each temperature.

\newpage

\begin{table}[t!]
	\caption{Number of beads used in RPMD rate calculations. $N_{\text{H/D}}$ refers to the number of beads used to simulate a hydrogen and a deuterium atom, respectively.
	\label{tab:rpmd-beads}}
	\begin{ruledtabular}
	\begin{tabular}{r*{6}{>{\centering\arraybackslash}m{4em}}}
		$T\, (\mathrm{K})$  & 50 & 60--70 & 80--100 & 125--150 & 160--225 & 250--300 \\
		\midrule
		\multirow{2}{*}{$N_{\text{H}}$} & 96 & 64 & 48 & 32 & 24 & 16  \\
					                    & 64 & 48 & 32 & 24 & 16 & 12 \\
		\midrule
		\multirow{2}{*}{$N_{\text{D}}$} & 64 & 48 & 32 & 24 & 16 & 12 \\
										& 48 & 32 & 24 & 16 & 12 & 8 \\
	\end{tabular}
	\end{ruledtabular}
\end{table}

\subsection{Convergence with spectral density cut-off}

\begin{figure}[h!]
	\includegraphics{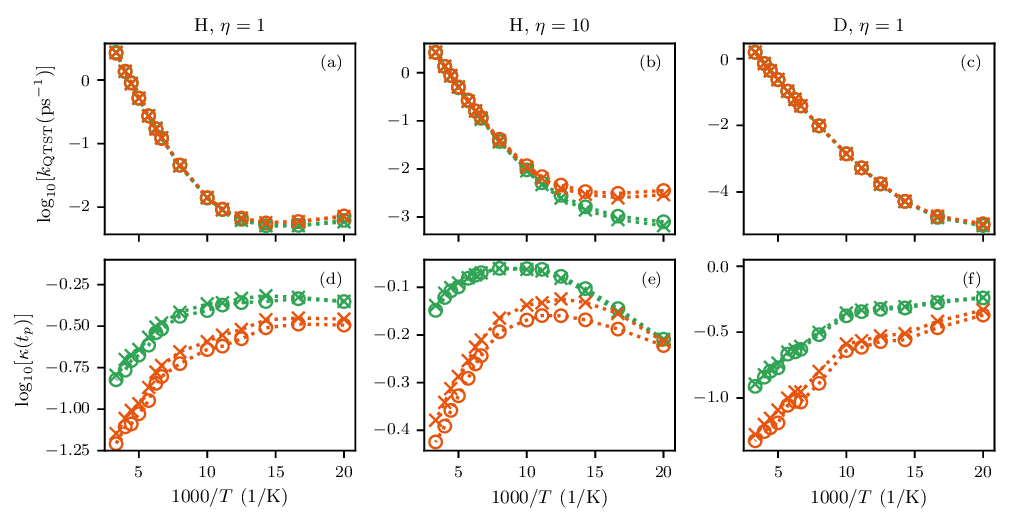}%
	\caption{Same as \fig{bead-conv}, except now the open circles refer to rates calculated for friction spectral densities damped by an exponential weight $w = \eu[-\omega/\omega_c]$ with a cut-off $\omega_c = \SI{2000}{\wn}$, and crosses denote rates for $\omega_c = \SI{4000}{\wn}$.
	\label{fig:wcut-conv}}
\end{figure}

To simplify the harmonic bath discretization~\cite{waltersDirectDetermination2017} and the fitting of auxiliary variable parameters (\sec{sec:auxvar-fit}), we performed all our calculations for memory--friction kernels with exponentially damped spectra densities, $\Lambda(\omega) \to \Lambda(\omega) \eu[-\omega/\omega_c]$. We tested two cut-off frequencies, $\omega_c = \SI{2000}{\wn}$ and $\SI{4000}{\wn}$, chosen to be large enough to have negligible impact on the calculated rates (see \fig{wcut-conv}). All calculations in the main article are quoted for $\omega_c = \SI{4000}{\wn}$.

\newpage

\subsection{Non-Markovian effects and the crossover temperature}

\begin{figure}[h!]
	\includegraphics{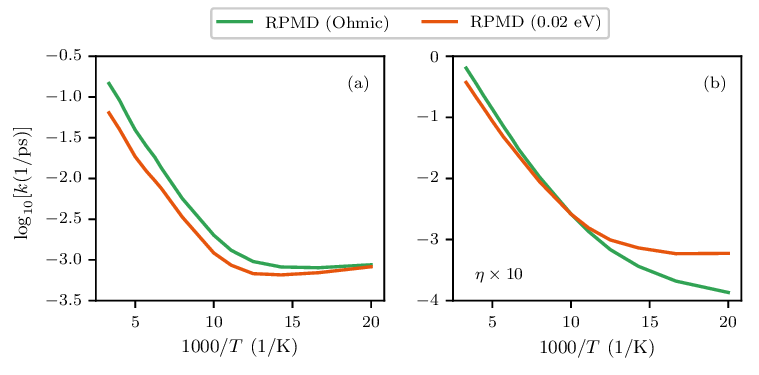}%
	\caption{%
	(a)~Hydrogen escape rates from the \emph{hcp} site of Cu(111) for the potential in Fig.~1(a), computed using the same methods and friction profiles as in Fig.~2(c). (b)~Same as (a), but for friction scaled by a factor of~10.
	\label{fig:cross}}
\end{figure}

In the Letter, we show that for weak damping, memory-friction effects are mostly manifested in the transmission coefficient, resulting in Ohmic and super-Ohmic rates that differ by an approximately constant scaling factor over a broad temperature range [\fig{cross}(a)].
As the damping strength increases, the activation free energy of the reaction also becomes influenced by the memory-friction effects.
The influence is most pronounced at low temperatures [Fig.~1(e)], where the Markovian approximation suppresses the rates. When combined with the changes to the dynamical transmission coefficient, this leads to a substantial shift in the tunnelling crossover temperature, as shown in \fig{cross}(b).

